\documentclass[10pt]{article}
\usepackage{amsmath,amsthm,latexsym,amssymb,amsfonts,epsfig, psfrag,color,dsfont}
\usepackage[utf8]{inputenc}
\usepackage{graphicx}
\usepackage{caption}
\usepackage{subcaption}
\usepackage{sidecap}
\usepackage{wrapfig}
\usepackage{url}
\usepackage{cite}
\usepackage{tikz-cd}
\usepackage{hyperref}
\usepackage{authblk}

\makeatletter \@addtoreset{equation}{section} \makeatother

\def\be{\begin{equation}}
\def\ee{\end{equation}}
\def\ba{\begin{eqnarray}}
\def\ea{\end{eqnarray}}

\newcommand\nn{\nonumber}
\newcommand\q{\quad}
\def\Nl{{\mathchoice
{\setbox0=\hbox{$\displaystyle\rm N$}\hbox{\hbox to0pt
{\kern0.4\wd0\vrule height0.9\ht0\hss}\box0}}
{\setbox0=\hbox{$\textstyle\rm N$}\hbox{\hbox to0pt
{\kern0.4\wd0\vrule height0.9\ht0\hss}\box0}}
{\setbox0=\hbox{$\scriptstyle\rm N$}\hbox{\hbox to0pt
{\kern0.4\wd0\vrule height0.9\ht0\hss}\box0}}
{\setbox0=\hbox{$\scriptscriptstyle\rm N$}\hbox{\hbox to0pt
{\kern0.4\wd0\vrule height0.9\ht0\hss}\box0}}}}
\def\Zl{{\mathchoice
{\setbox0=\hbox{$\displaystyle\rm Z$}\hbox{\hbox to0pt
{\kern0.4\wd0\vrule height0.9\ht0\hss}\box0}}
{\setbox0=\hbox{$\textstyle\rm Z$}\hbox{\hbox to0pt
{\kern0.4\wd0\vrule height0.9\ht0\hss}\box0}}
{\setbox0=\hbox{$\scriptstyle\rm Z$}\hbox{\hbox to0pt
{\kern0.4\wd0\vrule height0.9\ht0\hss}\box0}}
{\setbox0=\hbox{$\scriptscriptstyle\rm Z$}\hbox{\hbox to0pt
{\kern0.4\wd0\vrule height0.9\ht0\hss}\box0}}}}
\def\Ql{{\mathchoice
{\setbox0=\hbox{$\displaystyle\rm Q$}\hbox{\hbox to0pt
{\kern0.4\wd0\vrule height0.9\ht0\hss}\box0}}
{\setbox0=\hbox{$\textstyle\rm Q$}\hbox{\hbox to0pt
{\kern0.4\wd0\vrule height0.9\ht0\hss}\box0}}
{\setbox0=\hbox{$\scriptstyle\rm Q$}\hbox{\hbox to0pt
{\kern0.4\wd0\vrule height0.9\ht0\hss}\box0}}
{\setbox0=\hbox{$\scriptscriptstyle\rm Q$}\hbox{\hbox to0pt
{\kern0.4\wd0\vrule height0.9\ht0\hss}\box0}}}}
\def\Rl{{\mathchoice
{\setbox0=\hbox{$\displaystyle\rm R$}\hbox{\hbox to0pt
{\kern0.4\wd0\vrule height0.9\ht0\hss}\box0}}
{\setbox0=\hbox{$\textstyle\rm R$}\hbox{\hbox to0pt
{\kern0.4\wd0\vrule height0.9\ht0\hss}\box0}}
{\setbox0=\hbox{$\scriptstyle\rm R$}\hbox{\hbox to0pt
{\kern0.4\wd0\vrule height0.9\ht0\hss}\box0}}
{\setbox0=\hbox{$\scriptscriptstyle\rm R$}\hbox{\hbox to0pt
{\kern0.4\wd0\vrule height0.9\ht0\hss}\box0}}}}
\def\Cl{{\mathchoice
{\setbox0=\hbox{$\displaystyle\rm C$}\hbox{\hbox to0pt
{\kern0.4\wd0\vrule height0.9\ht0\hss}\box0}}
{\setbox0=\hbox{$\textstyle\rm C$}\hbox{\hbox to0pt
{\kern0.4\wd0\vrule height0.9\ht0\hss}\box0}}
{\setbox0=\hbox{$\scriptstyle\rm C$}\hbox{\hbox to0pt
{\kern0.4\wd0\vrule height0.9\ht0\hss}\box0}}
{\setbox0=\hbox{$\scriptscriptstyle\rm C$}\hbox{\hbox to0pt
{\kern0.4\wd0\vrule height0.9\ht0\hss}\box0}}}}
\def\Hl{{\mathchoice
{\setbox0=\hbox{$\displaystyle\rm H$}\hbox{\hbox to0pt
{\kern0.4\wd0\vrule height0.9\ht0\hss}\box0}}
{\setbox0=\hbox{$\textstyle\rm H$}\hbox{\hbox to0pt
{\kern0.4\wd0\vrule height0.9\ht0\hss}\box0}}
{\setbox0=\hbox{$\scriptstyle\rm H$}\hbox{\hbox to0pt
{\kern0.4\wd0\vrule height0.9\ht0\hss}\box0}}
{\setbox0=\hbox{$\scriptscriptstyle\rm H$}\hbox{\hbox to0pt
{\kern0.4\wd0\vrule height0.9\ht0\hss}\box0}}}}
\def\Ol{{\mathchoice
{\setbox0=\hbox{$\displaystyle\rm O$}\hbox{\hbox to0pt
{\kern0.4\wd0\vrule height0.9\ht0\hss}\box0}}
{\setbox0=\hbox{$\textstyle\rm O$}\hbox{\hbox to0pt
{\kern0.4\wd0\vrule height0.9\ht0\hss}\box0}}
{\setbox0=\hbox{$\scriptstyle\rm O$}\hbox{\hbox to0pt
{\kern0.4\wd0\vrule height0.9\ht0\hss}\box0}}
{\setbox0=\hbox{$\scriptscriptstyle\rm O$}\hbox{\hbox to0pt
{\kern0.4\wd0\vrule height0.9\ht0\hss}\box0}}}}
\newcommand{\cc}{\mathcal C}

\newcommand{\cg}{\mathcal G}
\newcommand{\ch}{\mathcal H}

\newcommand{\cp}{\mathcal P}

\newcommand{\cs}{\mathcal S}
\newcommand{\ct}{\mathcal T}

\def\nn{\nonumber}

\newcommand{\eqa}{\begin{eqnarray}}
\newcommand{\neqa}{\end{eqnarray}}

\def\la{\langle}
\def\ra{\rangle}
\newcommand{\bra}[1]{\la {#1}|}
\newcommand{\ket}[1]{|{#1}\ra}

\newcommand{\p}{\partial}

\def\f{\frac}

\usepackage{bbm}

\def\q{{\quad}}

\definecolor{bianca}{rgb}{0,0.,0.8}

\title{Switching internal times and a new perspective on the `wave function of the universe'}

\author[1,2]{Philipp A.\ H\"ohn\thanks{\texttt{p.hoehn@univie.ac.at}} }
\affil[1]{\small Institute for Quantum Optics and Quantum Information, Austrian Academy of Sciences,\newline Boltzmanngasse 3, 1090 Vienna, Austria}
\affil[2]{\small Vienna Center for Quantum Science and Technology (VCQ), Faculty of Physics, University of Vienna, Boltzmanngasse 5, 1090 Vienna, Austria}

\date{}

\begin{document}

\maketitle

\vspace{-.9cm}

\begin{abstract}

Despite its importance in general relativity, a quantum notion of general covariance has not yet been established in quantum gravity and  cosmology, where, given the a priori absence of coordinates, it is necessary to replace classical  frames with dynamical quantum reference systems. 
As such, quantum general covariance bears on the ability to consistently switch between the descriptions of the {\it same} physics relative to arbitrary choices of quantum reference system. Recently, a systematic approach for such switches has been developed \cite{Vanrietvelde:2018pgb, Vanrietvelde:2018dit,Hoehn:2018aqt}. It  links the descriptions relative to different choices of quantum reference system, identified as the correspondingly reduced quantum theories, via the reference-system-neutral Dirac quantization, in analogy to coordinate changes on a manifold. In this work, we apply this method to a simple cosmological model to demonstrate how to consistently switch between different internal time choices in quantum cosmology. We substantiate the argument that the conjunction of Dirac and reduced quantized versions of the theory defines a complete relational quantum theory that not only admits a quantum general covariance, but, we argue, also suggests a new perspective on the `wave function of the universe'. It assumes the role of a perspective-neutral global state, without immediate physical interpretation, that, however, encodes all the descriptions of the universe relative to all possible choices of reference system at once and constitutes the crucial link between these internal perspectives. While, for simplicity, we use the Wheeler-DeWitt formulation, the method and arguments might be also adaptable to loop quantum cosmology.

\end{abstract}

\section{Introduction}

General covariance is a celebrated feature of general relativity. It asserts that all the laws of physics are the same in all reference frames and independent of coordinates. It not only permits us to describe the physics from arbitrary choices of reference frame, but also to switch between the different descriptions at will. General covariance is the origin of the diffeomorphism invariance of the theory and thereby leads to profound conceptual consequences \cite{Rovelli:2004tv}: physical systems are neither localized nor evolve with respect to a background spacetime, but relative to one another. General covariance thus already implies classically that coordinates are not a fundamental concept in physics. While they are practical for any concrete calculations of the physics {\it in} a given spacetime, already classically, one could, instead, use dynamical degrees of freedom as reference systems relative to which to describe the physics, incl.\ the dynamics {\it of} spacetime \cite{Rovelli:2004tv,Rovelli:1990ph,Rovelli:1990pi,Brown:1994py,Dittrich:2004cb, Dittrich:2005kc, Dittrich:2006ee,Dittrich:2007jx,Tambornino:2011vg}.

In quantum cosmology and quantum gravity the situation becomes more extreme: since one does not quantize spacetime and its matter content relative to a background, coordinate systems are a priori absent altogether. Consequently, it becomes a {\it necessity} to employ dynamical degrees of freedom as {\it quantum} reference systems relative to which to describe the physics \cite{Rovelli:2004tv,Rovelli:1990ph,Rovelli:1990pi,Brown:1994py,DeWitt:1967yk,Kuchar:1991qf,Isham:1992ms,Anderson:2017jij,Rovelli:1990jm, Rovelli:1989jn,Tambornino:2011vg,Marolf:1994nz, Marolf:2009wp,Dittrich:2016hvj, Dittrich:2015vfa ,Bojowald:2010xp,Bojowald:2010qw, Hohn:2011us, Thiemann:2007zz,martinbuch,Ashtekar:2011ni,Banerjee:2011qu,Oriti:2016qtz, Gielen:2016uft,Gielen:2018fqv, Blyth:1975is,Hawking:1983hn,Hajicek:1986ky,Kiefer:1988tr, Ashtekar:2007em, Bojowald:2011zzb,Ashtekar:2006uz,Kamenshchik:2018pwf }. Ordinary coordinate systems are only expected to be reconstructed from such reference systems in a semiclassical, large-scale limit.

A question that has so far received little attention in quantum gravity and cosmology is how to establish a {\it quantum} notion of general covariance, despite its fundamental importance to the theory supposed to be quantized. A reason is perhaps the absence of coordinates and the (attempted) outright diffeomorphism invariance in quantum gravity. However, already classically, general covariance is less about coordinates and, operationally, primarily about linking the descriptions relative to different reference frames. 
Similarly, given the absence of coordinates, quantum general covariance can only refer to the ability to consistently switch between the descriptions of the {same} physics relative to arbitrary choices of quantum reference system. This includes both spatial and temporal reference systems.

As an initial step, we shall address this question in the context of a simple isotropic and homogeneous quantum cosmological model in this article, exploiting a novel  framework for quantum reference systems~\cite{Vanrietvelde:2018pgb,Vanrietvelde:2018dit,Hoehn:2018aqt,Giacomini:2017zju} and building up on the earlier works \cite{Bojowald:2010xp,Bojowald:2010qw,Hohn:2011us}.
As such, we will here not be concerned with spatial reference systems \cite{Vanrietvelde:2018pgb, Vanrietvelde:2018dit,Giacomini:2017zju}, but only internal times to which one usually resorts for defining temporal localization in quantum cosmology \cite{DeWitt:1967yk,Blyth:1975is,Hawking:1983hn,Hajicek:1986ky,Kiefer:1988tr,Kuchar:1991qf,Isham:1992ms,Tambornino:2011vg,Marolf:1994nz, Marolf:2009wp,Dittrich:2016hvj,Dittrich:2015vfa,Bojowald:2010xp,Bojowald:2010qw, Hohn:2011us, Thiemann:2007zz,martinbuch,Ashtekar:2011ni,Banerjee:2011qu,  Ashtekar:2007em, Bojowald:2011zzb,Ashtekar:2006uz,Kamenshchik:2018pwf,Oriti:2016qtz, Gielen:2016uft,Gielen:2018fqv }.

The use of different choices of internal times in parametrized systems and cosmological models has been considered, e.g., in \cite{Blyth:1975is,Kuchar:1991qf,Hartle:1995vj,Hajicek:1999ti,Gambini:2000ht}, but no explicit switches between the different choices were constructed. Instead, the so-called {\it multiple choice problem} associated to the problem of time was diagnosed \cite{Kuchar:1991qf,Isham:1992ms}. This is the purported problem that generically there are no distinguished internal time choices and that different choices of internal times  would lead to unitarily inequivalent quantum theories. Switching between different internal time choices was only later studied in a semiclassical approach \cite{Bojowald:2010xp,Bojowald:2010qw, Hohn:2011us, Bojowald:2016fac} and, for a restricted set of choices, at the level of reduced quantization \cite{Malkiewicz:2014fja,Malkiewicz:2015fqa,Malkiewicz:2016hjr,Malkiewicz:2017cuw}. Nevertheless, the meaning of quantum general covariance remained elusive.

One of our aims here will be to begin clarifying both technically and conceptually what quantum general covariance is, at least in the simplified context of quantum cosmology. The method and concepts, however, extend, at least in principle, to full quantum gravity. To this end, I will invoke a recent unifying approach to switching quantum reference systems in both quantum foundations and gravity \cite{Vanrietvelde:2018pgb, Vanrietvelde:2018dit,Hoehn:2018aqt }. This approach blends operational quantum reference frame methods \cite{Giacomini:2017zju}, aiming at quantum covariance too, with the ideas underlying the semiclassical clock switches in \cite{Bojowald:2010xp,Bojowald:2010qw, Hohn:2011us, Bojowald:2016fac} and conceptual arguments concerning the `wave function of the universe' and how to accommodate different frame perspectives in it \cite{Hoehn:2017gst}. In particular, in \cite{Hoehn:2018aqt } it was already shown that it provides a systematic method for switching between different choices of relational quantum clocks and this will be exploited below.

The key feature of the method in \cite{Vanrietvelde:2018pgb, Vanrietvelde:2018dit,Hoehn:2018aqt } is that it identifies a consistent quantum reduction procedure that maps the Dirac quantized theory to the various reduced quantized versions of it relative to different choices of quantum reference systems. It identifies the physical Hilbert space of the Dirac quantization as a reference-system-neutral quantum super structure and the various reduced quantum theories as the physics described relative to the corresponding choice of reference system. In analogy to a coordinate change on a manifold, one can then switch between different choices of quantum reference system by inverting a given quantum reduction map and concatenating it with the forward reduction map associated to the new choice of reference system. Just like coordinate changes, this will not always work globally, but this, I argue, is the structure defining {\it quantum} general covariance in a canonical formulation \cite{Vanrietvelde:2018pgb, Vanrietvelde:2018dit,Hoehn:2018aqt }. In particular, a {\it complete} relational quantum theory, admitting quantum general covariance, is the conjunction of its Dirac and various reduced quantized versions, just like the classical theory contains both the constraint surface and reduced phase spaces \cite{Hoehn:2018aqt}. By linking the various (generally unitarily inequivalent) reduced quantizations, the multiple choice problem becomes a {\it multiple choice feature} of the complete relational quantum theory \cite{Hoehn:2018aqt}, just like general covariance is a feature of general relativity.

In this work, I will apply this method to the simple flat Friedman-Robertson-Walker (FRW) universe filled with a massless, homogeneous scalar field and show how to consistently switch between choosing either the scale factor or the field as an internal time in both the classical and quantum theory and how the different descriptions are explicitly linked. This model has become a fairly standard example in Wheeler-DeWitt type quantum cosmology \cite{Blyth:1975is,Hawking:1983hn,Kamenshchik:2018pwf} and loop quantum cosmology \cite{Ashtekar:2006uz,Ashtekar:2007em,Ashtekar:2011ni,Banerjee:2011qu} and has recently even been reconstructed from a full quantum gravity theory \cite{Oriti:2016qtz, Gielen:2016uft}. Our discussion will be of relevance to each of these approaches, although loop quantization related subtleties need be taken into account before this framework can be directly applied to the latter two approaches (see later comments).


In the conclusions, I will use this explicit construction to argue more generally that quantum general covariance also entails a novel perspective on the `wave function of the universe'. It provides the handle to relate quantum states of subsystems `as seen' by other subsystems to `the wave function of the universe', linking frame-dependent and frame-independent descriptions of the physics and thereby suggesting a new interpretation of states in quantum cosmology.  In particular, I propose to view the `wave function of the universe' as a perspective-neutral global state that does not admit an immediate physical interpretation, but that encodes all the descriptions of the universe relative to all possible choices of reference system at once and constitutes the crucial link between all these internal perspectives. This will substantiate (and partially amend) an earlier proposal for interpreting the `wave function of the universe' and rendering it compatible with operationally significant relative states \cite{Hoehn:2017gst} (see also the earlier discussion in \cite{Hoehn:2014uua}).

\section{The flat FRW model with massless scalar field}\label{sec_cl}

Consider an isotropic and homogeneous FRW universe, filled with a homogeneous scalar field $\phi(t)$ and described by a metric $\mathrm{d}s^2=-\mathrm{d}t^2+a^2(t)(\mathrm{d}r^2/(1-kr^2)+r^2\mathrm{d}\Omega^2)$, where $a(t)$ is the scale factor and $k=-1,0,+1$ characterize open, flat and closed universes, respectively. For quantization later, it will be convenient to rather choose $\alpha:=\ln a$, so that $(\alpha,\phi)\in\mathbb{R}^2$ will be our configuration variables. This choice also simplifies the form of the Hamiltonian constraint, generating the dynamics, and yields\footnote{In fact, we have included a choice of lapse function $N=e^{3\alpha}$.} $C_H=p_\phi^2-p_\alpha^2-4k\,\exp(4\alpha)+4m^2\,\phi^2\,\exp(6\alpha)$, where $m$ is the mass of the field, e.g.\ see \cite{Blyth:1975is,Hawking:1983hn,Hajicek:1986ky,Kiefer:1988tr,Ashtekar:2011ni,Ashtekar:2007em, Bojowald:2011zzb,Banerjee:2011qu,Ashtekar:2006uz,Kamenshchik:2018pwf}. For illustrative purposes, we shall henceforth set the mass and curvature to zero, $m=k=0$, such that the Hamiltonian constraint takes a particularly simple Klein-Gordon form
\ba
C_H=p_\phi^2-p_\alpha^2\approx0\,,\label{CH}
\ea
where $\approx$ denotes a weak equality \cite{Dirac,Henneaux:1992ig}. Hence, we can equivalently interpret the dynamics as either a flat FRW model with massless scalar field, or as a relativistic particle in 1+1 dimensions. 

To understand the quantum internal time switches, it is necessary to first carefully revisit the classical model.

\subsection{Classical relational dynamics and internal time switches}
 
It is clear that $p_\phi,p_\alpha$ are {\it dependent} constants of motion and thus Dirac observables, as are
\ba
 \Lambda=p_\alpha\,\phi+p_\phi\,\alpha\,,\q\q\q\q\q L=p_\phi\,\phi+p_\alpha\,\alpha.\label{L}
 \ea
We have not yet selected a temporal reference with respect to which to interpret the dynamics. The constraint surface $\cc$ defined by (\ref{CH}) encodes all possible internal time choices at once, as reflected also in the redundancy of its description, and constitutes an internal-time-neutral super structure \cite{Hoehn:2018aqt} (see also \cite{Vanrietvelde:2018pgb, Vanrietvelde:2018dit}). As such, $\cc$ itself does not admit the interpretation as the physics described relative to a reference system; it is also not a phase space, but a pre-symplectic manifold.

 Using $\Lambda$ or $L$, we can construct {\it relational} Dirac observables 
 \cite{Rovelli:1990jm, Rovelli:1989jn,Rovelli:1990pi, Rovelli:2004tv, Dittrich:2004cb, Dittrich:2005kc, Dittrich:2006ee,Dittrich:2007jx,Tambornino:2011vg,Dittrich:2016hvj, Dittrich:2015vfa ,Bojowald:2010xp,Bojowald:2010qw, Hohn:2011us, Hoehn:2018aqt,Marolf:1994nz, Marolf:2009wp } in various ways. For simplicity, we choose $\alpha,\phi$ as internal times, exploiting that they are globally monotonic. For compactness of notation, denote by $e$ and $t$ the evolving and clock configuration degree of freedom, respectively, which are either $e=\phi$ and $t=\alpha$, or vice versa. The relational observable describing the evolution of $e$ with respect to $t$ can be easily constructed by evaluating the right hand side of $\Lambda=p_\alpha\,\phi+p_\phi\,\alpha$ along the trajectories generated by $C_H$ (with flow parameter $s$) and noting that $\Lambda$ is a constant of motion, producing
 \ba
 e(\tau):=e(s)\left|_{t(s)=\tau}\right. =\f{1}{p_t}\,(\Lambda-p_e\,\tau)=-\f{p_e}{p_t}(\tau-t)+e\,.
 \ea 
(The situation is completely symmetric in $\alpha$ and $\phi$.) This parameter family of Dirac observables gives the value of $e$ when the clock $t$ reads $\tau$. We would have to carefully regularize the inverse powers of $p_t$ in the subsequent reduced phase spaces and quantum theory. While this can be done \cite{Hoehn:2018aqt}, it will be convenient to make a variable change in the evolving degrees of freedom to avoid these complications. Instead of the canonical pair $(e,p_e)$, we will henceforth look at the evolution of the {\it affine} pair $(E:=e\,p_e,p_e)$, satisfying $\{E,p_e\}=p_e$, with respect to $t$. 
This amounts to evaluating $L$ instead of $\Lambda$ and yields
  \ba
  E(\tau):=E(s)\left|_{t(s)=\tau}\right. \approx L-p_t\,\tau=-p_t\,(\tau-t)+E\,,\q\q\q\q  p_e(\tau):=p_e(s)\left|_{t(s)=\tau}\right. = p_e\,,\label{Dirac}  
 \ea 
so that we have no singular behavior to worry about. 
 
\begin{SCfigure}
\centering
		\includegraphics[scale=.6]{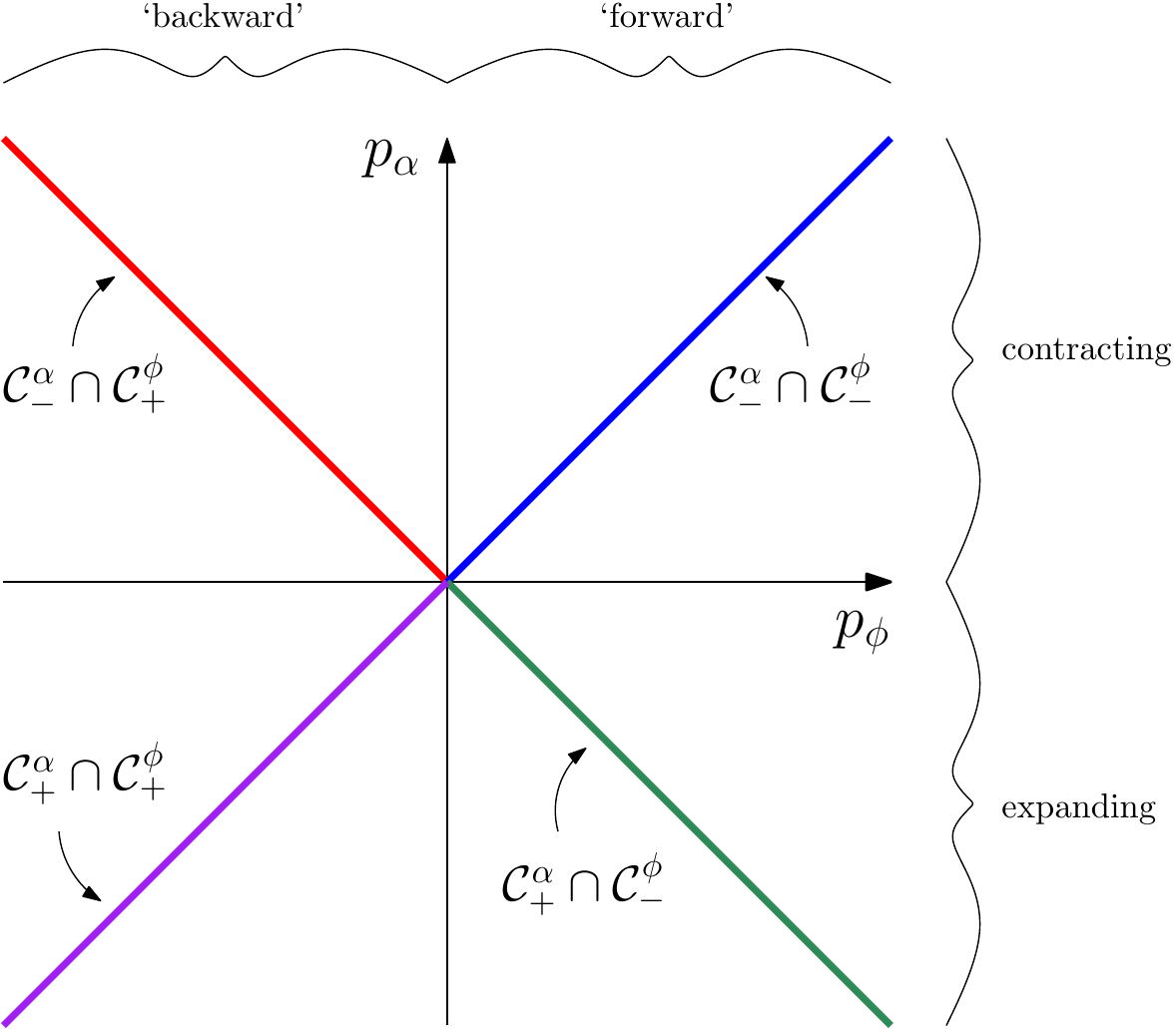}
		{\small{\caption{\label{fig:cs}\small  Schematic representation of the constraint surface $\cc$, defined by (\ref{CH}), as a `light cone' in momentum space. Its four components have the following physical interpretation. \\Red: contracting universe, but $\phi$ runs `backward'. \\Blue: contracting universe and $\phi$ runs `forward'. \\Green: expanding universe and $\phi$ runs `forward'. \\Purple: expanding universe, but $\phi$ runs `backward'. At the intersection point (the origin) the dynamics is static.}}}
\end{SCfigure}

We wish to remove the redundant clock degrees of freedom from among the dynamical variables through reduction \cite{Hoehn:2018aqt}. 
To this end, it will be convenient to factorize (\ref{CH}),
\ba
C_H=s_t\,C^t_+C^t_-\,,\q\q\q\q\q C^t_\pm:=p_t\pm h_e\,,\q\q\q\q\q h_e:=|p_e|\,,\q\q\q\q s_t:=\begin{cases}
  +1\,,    & t=\phi, \\
   -1\,,   & t=\alpha.
\end{cases}\label{cc}
\ea
$h_e$ will assume the role of a Hamiltonian. We have the following situation:

%
%

\begin{itemize}
\item[(i)] On $\cc^t_\pm\subset\cc$, defined by $C^t_\pm=0$ and $p_t\neq0$, we have
\ba
\f{\mathrm{d}\,\cdot}{\mathrm{d}s}=\{\cdot,C_H\}\approx \mp\,2s_t\,h_e\,\{\cdot,C^t_\pm\}\,,
\ea
so that $C^t_\pm$ generates the dynamics on $\cc^t_\pm$. Since $h_e>0$, the flows generated by $C^t_+$ and $C^t_-$ are opposite to and aligned with that of $C_H$, respectively, for $t=\phi$ and aligned with and opposite to that of $C_H$, respectively, for $t=\alpha$. That is, $\phi$ runs `backward' on $\cc^\phi_+$ and `forward' on $\cc^\phi_-$, while $\alpha$ expands on $\cc^\alpha_+$ and contracts on $\cc^\alpha_-$. Backward/expanding and forward/contracting will correspond to positive and negative frequency solutions, respectively, in the quantum theory.

\item[(ii)] The set $p_\alpha=p_\phi=0$ is the shared boundary between $\cc^\phi_+$ and $\cc^\phi_-$, as well as between $\cc^\alpha_+$ and $\cc^\alpha_-$. Notice that orbits with $p_\alpha=p_\phi=0$ are just points in $\cc$ so that the latter is stratified by gauge orbits of different dimension. 
Since $\mathrm{d}C_H=0$ for $p_\alpha=p_\phi=0$, 
no gauge-fixing surface can pierce every such gauge orbit once and only once. 

\end{itemize}
The situation is summarized in fig.\ \ref{fig:cs} for convenience.

On $\cc^t_\pm$ we can thus use $C^t_\pm$ as evolution generators and their {\it relational} dynamics is equivalent to that of $C_H$. Indeed, on $\cc^t_\pm$ we find ($s_\pm$ denotes the flow parameter of $C^t_\pm$):
\ba
E_\pm(\tau):=E(s_\pm)\left|_{t(s_\pm)=\tau}\right. =\pm\,|p_e|\,(\tau-t)+E\,,\q\q\q\q  p_{e\pm}(\tau):=p_e(s_\pm)\left|_{t(s_\pm)=\tau}\right. = p_e\,,\label{Dirac2}
 \ea 
 which is (\ref{Dirac}) after solving (\ref{CH}). 
 The relational Dirac observables being gauge-invariant extensions of gauge-restricted quantities \cite{Dittrich:2004cb, Dittrich:2005kc,Henneaux:1992ig,Hoehn:2018aqt}, we can now gauge fix the clock to, e.g., $t=0$  and evaluate (\ref{Dirac2}) on this surface $\cg_{t=0}$ without loss of dynamical information. This will produce two {\it separate} reduced phase spaces $\cp_\pm^{e(t)}\simeq\bar{\cc}^t_\pm\cap\cg_{t=0}$ for positive/negative frequency modes, where $\bar{\cc}^t_\pm$ is $\cc^t_\pm$ including its boundary $p_\alpha=p_\phi=0$. Of course, due to (ii) these gauge-fixed reduced phase spaces will miss all point-like orbits with $p_\alpha=p_\phi=0$ and $t\neq0$ and so their union does {\it not} coincide with the space of orbits $\cc/\sim$, where $\sim$ identifies points in the same orbit. We comment on this shortly. 

The Dirac bracket for any functions $F,G$ on $\cc^t_\pm$ reads
\ba
\{F,G\}_{D_\pm}=\{F,G\}-\{F,C^t_\pm\}\{t,G\}+\{F,t\}\{G,C^t_\pm\}\,.\label{Dpm}
\ea
All Dirac brackets involving the redundant clock variables $(t,p_t)$ vanish, which can thus be removed. Furthermore, the affine bracket
\ba
 \{E,p_e\}_{D_\pm}\equiv\{E,p_e\}=p_e\,\label{aff}
\ea
is well-defined everywhere. By contrast, the canonical $\{e,p_e\}_{D_\pm}$ is undefined for $p_e=0$. Hence, we take $\cp^{e(t)}_\pm$ to be fundamentally defined through the affine algebra (\ref{aff}). Then we could {\it define} $e:=E/p_e$ on $\cp^{e(t)}_\pm$, yielding a {\it derived} canonical relation $\{e,p_e\}_{D_\pm}=1$.

On $\cp^{e(t)}_\pm$ the relational observables (\ref{Dirac2}) become
\ba
E_\pm(\tau)=\pm\,|p_e|\,\tau+E\,,\q\q\q\q  p_{e\pm}(\tau) = p_e\,,\label{Dirac3}
 \ea 
and satisfy the following equations of motion
\ba
\f{\mathrm{d}E_\pm}{\mathrm{d}\tau}=\pm\,|p_e|=\{E_\pm,\pm \,h_e\}_{D_\pm}\,\q\q\q\q\q \f{\mathrm{d}p_{e_\pm}}{\mathrm{d}\tau}=0=\{p_{e_\pm},\pm \,h_e\}_{D_\pm}\,,
\ea
which are thus generated by the physical Hamiltonian $\pm\,h_e$. 

This can now genuinely be interpreted as the evolution described relative to the clock $t$, which, being the reference system, has become dynamically redundant and an evolution {\it parameter} $\tau$ (see also \cite{Hoehn:2018aqt}). Notice also that the measure-zero set of ignored orbits that distinguishes (the union of) $\cp^{e(t)}_\pm$ from the space of orbits is redundant for relational dynamics. Indeed, the ignored orbits correspond to the static point-like orbits with $p_\alpha=p_\phi=0$ and $t\neq0$, where $(E,p_e)$ are directly independent observables. But all their information is already encoded in (\ref{Dirac3}): for $p_e=0$, $E_\pm(\tau)=E$ does not depend on $\tau$, which, however, runs over all possible values of $t$. It is thus physically justified to work with the gauge-fixed reduced phase space $\cp^{e(t)}_\pm$ rather than the abstract reduced phase space $\cc/\sim$.
We will also see that the relation between the Dirac and reduced quantized theories is consistent with this observation.

Next, we interchange the roles of $e$ and $t$, i.e.\ we switch to using $e$ as the clock and $t$ as an evolving variable \cite{Hoehn:2018aqt}. The corresponding map between the corresponding reduced phase spaces $\cp^{e(t)}_\pm$ and $\cp_\pm^{t(e)}$ involves the gauge transformation generated by $C_H$ which maps $\cc\cap\cg_{t=0}$ to, e.g., $\cc\cap\cg_{e=0}$. Solving the equations of motion generated by $C_H$, one easily finds that one has to flow a parameter distance $s=s_t\,t_0/2p_t$ in $\cc$, where $t_0$ is the clock value prior to the transformation. Dropping the redundant variables, this yields the following maps\footnote{For more details of this procedure in a different model, see~\cite{Hoehn:2018aqt}.}
\ba
\cs_{t_+\to e_\pm}:\cp^{e(t)}_+\rightarrow\cp^{t(e)}_\pm\,,\q\q\q\q\q (E,\,p_e)\mapsto (T=E,\,\,p_t=-|p_e|)\,,\nn\\
\!\!\!\!\cs_{t_-\to e_\pm}:\cp^{e(t)}_-\rightarrow\cp^{t(e)}_\pm\,,\q\q\q\q\q (E,\,p_e)\mapsto (T=E,\,\,p_t=+|p_e|)\,,\label{clswitch}
\ea
where $T=t\,p_t$ is the evolving affine variable after the clock switch. Notice that gauge transformations preserve $\cc^\alpha_i\cap\cc^\phi_j$, $i,j=+,-$, i.e.\ the four quadrants of fig.\ \ref{fig:cs}. Hence, e.g., $\cs_{t_+\to e_\pm}$ maps the $p_e<0$ and $p_e>0$ halfs of $\cp^{e(t)}_+$ onto the $p_t<0$ halfs of $\cp^{t(e)}_+$ and $\cp^{t(e)}_-$, respectively, etc. (For example, $\cs_{\alpha_+\to\phi_-}$ switches from the description of the `expanding-forward' sector (green quadrant in fig.\ \ref{fig:cs}) relative to $\alpha$ to its description relative to $\phi$.) Respecting this, one obtains a `continuous' relational evolution, despite the clock switch: Using (\ref{Dirac3}) and setting $\tau^i_e=E_+(\tau_t^f)/p_e= :e_+(\tau_t^f)$ as the initial value of the new clock $e$ after the clock switch, where $\tau_t^f$ was the final value of the old clock $t$ prior to it, one consistently finds\footnote{Note that generally $T_\pm(\tau_e^i)\neq E_+(\tau_t^f)$, despite the form (\ref{clswitch}).} 
\ba
T_\pm(\tau^i_e)=p_t\,\tau_t^f\,,\q\q\q\text{ on }\cp^{t(e)}_\pm\,.\label{ccontinuous}
\ea
We shall see the quantum analog of this later.
We emphasize that due to the intermediate gauge transformation the clock switch proceeds via the internal-time-neutral $\cc$ \cite{Hoehn:2018aqt}.

%

\subsection{Reduced quantization relative to a choice of internal time}

We proceed by quantizing the gauge-fixed reduced phase spaces $\cp_\pm^{e(t)}$ of this model universe. Subsequently, we will link the various reduced quantum theories {\it via} the internal-time-neutral Dirac quantized theory. For simplicity, we resort to the Wheeler-DeWitt formulation in the Dirac procedure, but we note that the loop quantization of this FRW model can be cast into a very similar form (modulo observables) \cite{Ashtekar:2007em}. There is thus good hope that the below framework for switching internal times can be adapted to loop quantum cosmology. Before doing so, however, one has to overcome loop quantization related subtleties, which I briefly comment on in the conclusions. 

\begin{figure}
\centering
    \begin{tikzcd}[row sep=large, column sep = large]
& & \textrm{kinematical phase space } \mathbb{R}^4 \arrow{lld}[swap]{C^\phi_+ = \phi =0} \arrow[ld, "\hspace{-.3cm}C^\phi_- =\phi =0  \hspace{2.2cm}    C^\alpha_+=\alpha=0    "] 
\arrow[d,"\textrm{\hspace{-.82cm}Dirac \,quant.}"]
\arrow{rd}
 \arrow[rrd, "C_-^\alpha = \alpha =0"] & & \\
\mathcal{P}^{\alpha(\phi)}_+ \arrow[d,"\textrm{      \hspace*{-.8cm}  red.\  quantization rel.\  to $\phi$}"] & \mathcal{P}^{\alpha(\phi)}_- \arrow{d} & \mathcal{H}_\textrm{kin} \arrow[dd,"\delta(\hat{C}_H)"]  & \mathcal{P}^{\phi(\alpha)}_+ \arrow[d,"\textrm{\hspace*{-.6cm}red.\  quantization rel.\ to $\alpha$}"] & \mathcal{P}_-^{\phi(\alpha)} \arrow{d} \\
\mathcal{H}^{\alpha(\phi)}_+ & \mathcal{H}^{\alpha(\phi)}_- & &  \mathcal{H}^{\phi(\alpha)}_+ & \mathcal{H}^{\phi(\alpha)}_-\\
& \mathcal{H}_\textrm{phys}^{\alpha(\phi)} \arrow[lu, "\sqrt{4\pi}\bra{\phi=0} \theta(- \hat{p}_\phi) \widehat{\sqrt{|p_\alpha|}}"] \arrow{u}[swap]{\sqrt{4\pi}\bra{\phi=0} \theta( \hat{p}_\phi) \widehat{\sqrt{|p_\alpha|}}} & \mathcal{H}_\textrm{phys} \arrow[l, "\mathcal{T}_\phi"] \arrow{r}[swap]{\mathcal{T}_\alpha} & \mathcal{H}_\textrm{phys}^{\phi(\alpha)} \arrow[u, "\sqrt{4\pi}\bra{\alpha=0} \theta(- \hat{p}_\alpha) \widehat{\sqrt{|p_\phi|}}"] \arrow{ru}[swap]{\sqrt{4\pi}\bra{\alpha=0} \theta(\hat{p}_\alpha) \widehat{\sqrt{|p_\phi|}}} &
\end{tikzcd}    
\caption{{\small Diagrammatic overview of the relation between Dirac and the four reduced quantizations. In a nutshell, the physical Hilbert space is mapped to any of the four (positive or negative frequency) reduced Hilbert spaces by first trivializing the Hamiltonian constraint via $\ct_\phi$ or $\ct_\alpha$ to the corresponding choice of internal time variable and subsequently projecting onto the classical internal time gauge fixing condition. (Details in the main text.) }} \label{fig:sum}
\end{figure}
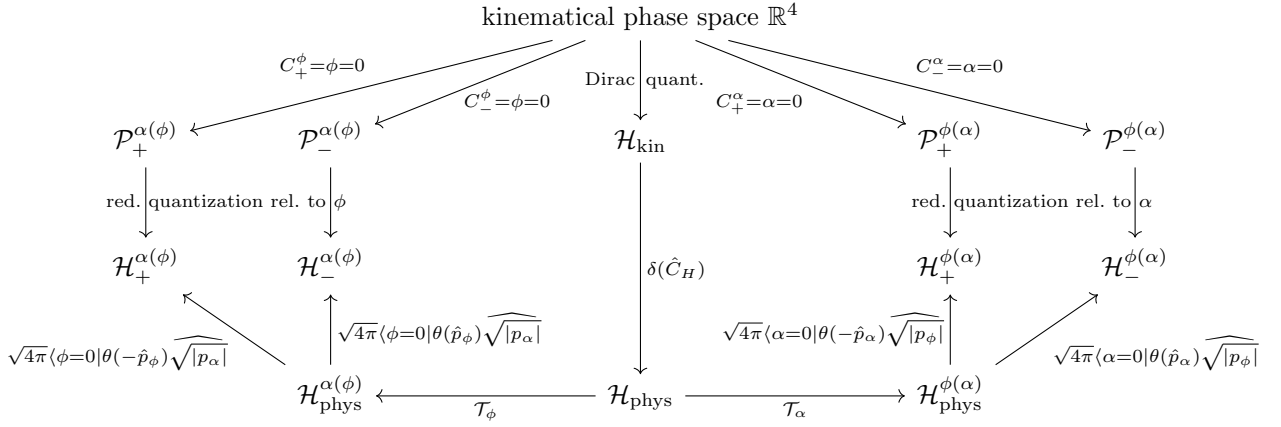

Since we will encounter a number of Hilbert spaces and transformations along the way, we summarize the various classical and quantum reduction steps and their relation in fig.\ \ref{fig:sum} for guidance.

Recall that $\cp^{e(t)}_\pm$ is defined through the affine algebra (\ref{aff}). However, it turns out to be equivalent to quantize these phase spaces in either the affine or standard canonical method. We promote the Dirac bracket $\{.,.\}_{D_\pm}$ to a commutator $[.,.]$ and $(e,p_e)$ to conjugate or $(E,p_e)$ to affinely related operators on a Hilbert space $\ch^{e(t)}_\pm:=L^2(\mathbb{R})$. In the canonical momentum representation, we represent states as
\ba
\ket{\psi}_\pm^{e(t)}=\int_{-\infty}^{+\infty}\,\mathrm{d}p_e\,\psi_\pm^{e(t)}(p_e)\,\ket{p_e}_e\,,\label{redstate}
\ea
the inner product as
\ba
\la\psi|\chi\ra^{e(t)}_\pm=\int_{-\infty}^{+\infty}\,\mathrm{d}p_e\,[\psi_\pm^{e(t)}(p_e)]^*\,\chi_\pm^{e(t)}(p_e)\,,\label{redIP}
\ea
$\hat{p}_e$ as a multiplication operator and the configuration observables as\footnote{We set $\hbar=1$.}
\ba
\hat{e}\,\psi^{e(t)}_\pm(p_e)= i\,\p_{p_e}\,\psi_\pm^{e(t)}(p_e)\,,\q\q\q\q \hat{E}\,\psi_\pm^{e(t)}(p_e)= i\,\left(p_e\,\p_{p_e}+\f{1}{2}\right)\,\psi_\pm^{e(t)}(p_e)\,.\label{affrep}
\ea
These are self-adjoint and for states with $\lim_{p_e\to\pm\infty}\,\sqrt{|p_e|}\,\psi_\pm^{e(t)}(p_e)=0$ we can equivalently work with $\hat{e}$ or $\hat{E}$ which also satisfies $\hat{E}=\f{1}{2}(\hat{e}\,\hat{p}_e+\hat{p}_e\,\hat{e})$.\footnote{In the affine momentum representation, states are represented as $\ket{\psi}_\pm^{e(t)}=\int_{-\infty}^{+\infty}\,\f{\mathrm{d}p_e}{|p_e|}\,\tilde{\psi}_\pm^{e(t)}(p_e)\,\ket{p_e}_{\rm aff}$, where $\tilde{\psi}_\pm^{e(t)}=\sqrt{|p_e|}\,\psi^{e(t)}_\pm$ and $\la p_e|p_e'\ra_{\rm aff}=|p_e|\,\delta(p_e-p_e')$. The inner product then reads $\la\psi|\chi\ra^{e(t)}_\pm=\int_{-\infty}^{+\infty}\,\f{\mathrm{d}p_e}{|p_e|}\,[\tilde\psi_\pm^{e(t)}(p_e)]^*\,\tilde\chi_\pm^{e(t)}(p_e)$ and the configuration observables are represented as $\hat{E}\,\tilde\psi_\pm^{e(t)}= i\,p_e\,\p_{p_e}\,\tilde\psi_\pm^{e(t)}$ and $\hat{e}\,\tilde\psi_\pm^{e(t)}=(i\,\p_{p_e}-\f{i}{2p_e})\,\tilde\psi_\pm^{e(t)}$. It is easy to check that this affine representation is equivalent to the canonical one above.} The evolving observables (\ref{Dirac3}) become
\ba
\hat{E}_\pm(\tau)=\pm\,|\hat{p}_e|\,\tau+\hat{E}\,,\q\q\q\q  \hat{p}_{e\pm}(\tau) = \hat{p}_e\,,\label{Dirac4}
 \ea
and satisfy the Heisenberg equations with Hamiltonian $\hat{H}=\pm\hat{h}_e=\pm|\hat{p}_e|$ on $\ch^{e(t)}_\pm$
\ba
\f{\mathrm{d}\hat{E}_\pm}{\mathrm{d}\tau}=\pm\,|\hat{p}_e|=-i\,[\hat{E}_\pm,\hat{H}]\,\q\q\q\q\q \f{\mathrm{d}\hat{p}_{e_\pm}}{\mathrm{d}\tau}=0=-i\,[\hat{p}_{e_\pm},\hat{H}]\,.
\ea

\subsection{The internal-time-neutral Dirac quantization}

We continue with Dirac quantization (see fig.\ \ref{fig:sum}), promoting $(\alpha,p_\alpha)$ and $(\phi,p_\phi)$ to conjugate operators on a kinematical Hilbert space $\ch_{\rm kin}:=L^2(\mathbb{R}^2)$. The solutions to the quantum constraint
\ba
\hat{C}_H\,\ket{\psi}_{\rm phys}=(\hat{p}_\phi^2-\hat{p}_\alpha^2)\,\ket{\psi}_{\rm phys}\overset{!}{=}0\,.\label{qCH}
\ea
will define the physical Hilbert space $\ch_{\rm phys}$. Using {\it group averaging} \cite{Hartle:1997dc,Marolf:1995cn,Marolf:2000iq,Thiemann:2007zz,Vanrietvelde:2018pgb, Vanrietvelde:2018dit,Hoehn:2018aqt,Ashtekar:1995zh}, $\ket{\psi}_{\rm phys}=\delta(\hat{C}_H)\,\ket{\psi}_{\rm kin}$, and working in momentum representation with kinematical wave functions $\psi_{\rm kin}(p_\phi,p_\alpha$), we find physical states to be of the form
\ba
\ket{\psi}_{\rm phys}=\int_{-\infty}^{+\infty}\,\f{\mathrm{d}p_e}{2|p_e|}\Big[\psi^{e(t)}_{\rm kin}\left(-|p_e|,p_e\right)\,\ket{-|p_e|\,}_t\ket{p_e}_e+\psi^{e(t)}_{\rm kin}\left(|p_e|,p_e\right)\,\ket{\,|p_e|\,}_t\ket{p_e}_e\Big]\,\label{physstate}
\ea
and the physical inner product as
\ba
\la\psi|\chi\ra_{\rm phys}=\int_{-\infty}^{+\infty}\,\f{\mathrm{d}p_e}{2|p_e|}\Big[\big(\psi^{e(t)}_{\rm kin}(-|p_e|,p_e)\big)^*\,\chi^{e(t)}_{\rm kin}(-|p_e|,p_e)+\big(\psi^{e(t)}_{\rm kin}(|p_e|,p_e)\big)^*\,\chi^{e(t)}_{\rm kin}(|p_e|,p_e)\Big]\,,\label{PIP}
\ea
where for compactness of notation we have set
\ba\label{compact}
\psi_{\rm kin}^{e(t)}\big(p_t=\mp|p_e|,p_e\big):=\begin{cases}
 \psi_{\rm kin}(\mp|p_\alpha|,p_\alpha)\,,     & t=\phi\,,\q e=\alpha\,, \\
    \psi_{\rm kin}(p_\phi,\mp|p_\phi|)\,,  & t=\alpha\,,\q e=\phi\,.
    \end{cases}
    \ea
 The situation is completely symmetric in $\alpha$ and $\phi$ and we will exploit this for the internal time switches.
For interpretation it is useful to note that  the position representation of the states reads
\ba
\psi_{\rm phys}^\pm(e,t)=\int_{-\infty}^{+\infty}\,\f{\mathrm{d}p_e}{4\pi\,|p_e|}\,e^{i(\mp|p_e|\,t+p_e\,e)}\,\psi^{e(t)}_{\rm kin}\left(\mp|p_e|,p_e\right)\,,
\ea
where $\psi^\pm_{\rm phys}$ are the positive/negative frequency solutions of the Klein-Gordon equation. It is easy to convince oneself that
\ba
\la\psi|\chi\ra_{\rm phys}=2\pi\Big[\left(\psi_{\rm phys}^+,\chi_{\rm phys}^+\right)_{\rm KG}-\left(\psi_{\rm phys}^-,\chi_{\rm phys}^-\right)_{\rm KG}\Big]\,,
\ea
where $(\psi,\chi)_{\rm KG}=i\,\int\,\mathrm{d}e\,(\psi^*\p_t\chi-(\p_t\psi^*)\chi)$ is the usual Klein-Gordon inner product in which positive and negative frequency solutions are orthogonal (see also \cite{Hartle:1997dc}). Physical states and inner product thus decompose into a sum of positive and negative frequency modes. It follows from fig.\ \ref{fig:cs} that for $e=\alpha$ and $t=\phi$ positive/negative frequency solutions correspond to classical backward/forward evolution in $\phi$. Conversely, for $e=\phi$ and $t=\alpha$, positive/negative frequency solutions correspond to evolving relative to an expanding/contracting $\alpha$.
 It is standard (and usually justified) to ignore the negative frequency solutions \cite{Ashtekar:2007em,Ashtekar:2011ni,Banerjee:2011qu}; here we shall not do that as they will be interesting when switching internal times. In particular, it is easy to convince oneself, using (\ref{PIP}) and fig.\ \ref{fig:cs}, that both the positive and negative frequency part of a physical state  for $e=\alpha$ overlap with both the positive and negative frequency part of the same physical state associated to $e=\phi$.

Choosing a symmetric ordering, the relational Dirac observables (\ref{Dirac}) are quantized as
\ba
\hat{E}(\tau)=-\hat{p}_t\,\tau+\f{1}{2}\Big(\hat{p}_t\,\hat{t}+\hat{t}\,\hat{p}_t+ \hat{e}\,\hat{p}_e+\hat{p}_e\,\hat{e} \Big)+i=-\hat{p}_t\,\tau+\hat{t}\,\hat{p}_t+ \hat{p}_e\,\hat{e}+i\,,\q\q\q\q  \hat{p}_e(\tau)= \hat{p}_e\,\label{Dirac5}  
\ea
and commute with $\hat{C}_H$, however, $\hat{E}(\tau)$ {\it only} does so on $\ch_{\rm phys}$. This is also the reason for the $+i$ term, which ensures that $\hat{E}(\tau)$ is Hermitian with respect to (\ref{PIP}) and ultimately self-adjoint on $\ch_{\rm phys}$, see appendix \ref{app_SA}.

In analogy to the classical $\cc$, I propose to conceive of $\ch_{\rm phys}$ as the internal-time-neutral quantum structure \cite{Hoehn:2018aqt}. In the Dirac quantized theory, we have not yet chosen a temporal reference system with respect to which to interpret the dynamics. This is reflected in the redundancy of the representation of states (\ref{physstate}), inner product (\ref{PIP}) and relational observables (\ref{Dirac5}); we have not yet decided whether $t=\alpha$ or $\phi$ and we could have selected a different internal time altogether. Just like $\cc$, $\ch_{\rm phys}$ encodes all internal clock choices at once and it features no Heisenberg evolution equations for relational observables.

\subsection{Quantum reduction: from Dirac to reduced quantization}

Next, we perform the quantum reduction procedure that maps the Dirac to the various reduced quantized theories \cite{Vanrietvelde:2018pgb, Vanrietvelde:2018dit,Hoehn:2018aqt} and ultimately permits us to switch internal times also in the quantum theory. In analogy to the classical case, it proceeds as follows (see fig.\ \ref{fig:sum}):
(i) choose an internal time, (ii) trivialize the constraint to the internal time to render it redundant, (iii) project onto the classical gauge fixing conditions, corresponding to the choice of internal time, to remove the redundancy.

We define the {\it trivialization map}
\ba
\ct_t:=\ct_{t+}+\ct_{t-}\,,\q\q\q\q \ct_{t\pm}:=\exp\left(\pm i\,\hat{t}\,(\hat{h}_e-\epsilon)\right)\,\theta(\mp\hat{p}_t)\,,\label{trivialt}
\ea
where $\theta(0)=\f{1}{2}$. The theta function separates positive and negative frequency modes and the transformation is akin to the time evolution map in $t$ time, except that the latter appears as an operator. In consequence, $\ct_t$ does {\it not} commute with $\hat{C}_H$ and maps $\ch_{\rm phys}$ to a {new} Hilbert space $\ch_{\rm phys}^{e(t)}:=\ct_t(\ch_{\rm phys})$. Using the tools of \cite{Hoehn:2018aqt}, one can check that its inverse  $\ct_t^{-1}:\ch_{\rm phys}^{e(t)}\rightarrow\ch_{\rm phys}$ is given by 
\ba\label{trivialtinv}
\ct_t^{-1}:=\ct^{-1}_{t+}+\ct^{-1}_{t-}\,,\q\q\q\q \ct^{-1}_{t\pm}:=\exp\left(\mp i\,\hat{t}\,(\hat{h}_e-\epsilon)\right)\,\theta(\mp\hat{p}_t)\,.
\ea
and  satisfies
$\ct_t^{-1}\,\ct_t = \theta(-\hat{p}_t)+\theta(\hat{p}_t) = \mathds{1}$ {\it only} on $\ch_{\rm phys}$ and {\it only} for $\epsilon>0$. The role of the parameter $\epsilon$ is thus to render (\ref{trivialt}) invertible.

The key property of (\ref{trivialt}) is that it trivializes $\hat{C}_H$ to the clock variables. More precisely, 
\ba
\ct_{t\pm}\,\hat{C}^t_\pm\,\ct_{t\pm}^{-1} = (\hat{p}_t\pm\epsilon)\,\theta(\mp\hat{p}_t)\,,\q\q\q\q \ct_{t\mp}\,\hat{C}^t_\pm\,\ct_{t\mp}^{-1} = (\hat{p}_t\pm2\hat{h}_e\mp\epsilon)\,\theta(\pm\hat{p}_t)\,,
\ea
and so $\ct_{t\pm}$ trivializes $\hat{C}^t_\pm$ from (\ref{cc}) in the positive/negative frequency sector such that it {\it only} acts on the clock variables upon transformation.  Together
\ba
\ct_q\,\hat{C}_H\,\ct_q^{-1}=s_t\,\left(\hat{p}_t-2\,\hat{h}_e+\epsilon\right)\left(\hat{p}_t+\epsilon\right)\,\theta(-\hat{p}_t)+s_t\left(\hat{p}_t+2\,\hat{h}_e-\epsilon\right)\left(\hat{p}_t-\epsilon\right)\,\theta(\hat{p}_t)\,.\label{trivialq2}
\ea
It is thus not surprising to find the states of $\ch^{e(t)}_{\rm phys}$ in the form
\ba
\ket{\psi}_{\rm phys}^{e(t)}:=\ct_t\,\ket{\psi}_{\rm phys}=\int_{-\infty}^{+\infty}\,\f{\mathrm{d}p_e}{2|p_e|}\Big[\psi^{e(t)}_{\rm kin}\left(-|p_e|,p_e\right)\,\ket{-\epsilon}_t\ket{p_e}_e+\psi^{e(t)}_{\rm kin}\left(|p_e|,p_e\right)\,\ket{\epsilon}_t\ket{p_e}_e\Big]\,.\label{physstate2}
\ea
Hence, apart from distinguishing the positive/negative frequency sectors, the clock-slot of the state has become redundant. It is easy to convince oneself that $\ct_t$ constitutes an isometry from $\ch_{\rm phys}$ to $\ch_{\rm phys}^{e(t)}$.
%

After a straightforward calculation one finds that the relational Dirac observables (\ref{Dirac5}) transform as follows to $\ch_{\rm phys}^{e(t)}$:
\ba
\ct_t\,\hat{E}(\tau)\,\ct_t^{-1}&=&\left(|\hat{p}_e|\,\tau+\hat{p}_e\,\hat{e}+i\right)\,\theta(-\hat{p}_t)+\left(-|\hat{p}_e|\,\tau+\hat{p}_e\,\hat{e}+i\right)\,\theta(\hat{p}_t)\,,\nn\\
\ct_t\,\hat{p}_e(\tau)\,\ct_t^{-1}&=&\hat{p}_e\,\theta(-\hat{p}_t)+\hat{p}_e\,\theta(\hat{p}_t)\,.\label{toatrans1}
\ea
On the respective positive/negative frequency sectors, these almost coincide with the reduced evolving observables (\ref{Dirac4}) on $\ch^{e(t)}_\pm$.

To complete the quantum reduction to $\ch^{e(t)}_\pm$ we note the following:
\ba
\la\psi|\chi\ra_{\rm phys}\equiv\f{1}{2}\,\la\psi\ket{\chi}^{e(t)}_++\f{1}{2}\,\la\psi\ket{\chi}^{e(t)}_-\,,\label{inp}
\ea
where $\la\psi\ket{\chi}^{e(t)}_\pm$ is the inner product (\ref{redIP}), provided 
\ba
\psi^{e(t)}_\pm(p_e):=\f{\psi^{e(t)}_{\rm kin}(\mp|p_e|,p_e)}{\sqrt{|p_e|}}\,.\label{identify2}
\ea
The reduced state is thereby essentially the Newton-Wigner wave function associated to the positive/negative frequency solutions of the constraint (\ref{qCH}). However, there is a small difference: usually, one restricts to positive frequency solutions in which case the Newton-Wigner wave function involves an additional factor $1/\sqrt{2}$ \cite{haag2012local}. This would here imply $\la\psi|\chi\ra_{\rm phys}\equiv\,\la\psi\ket{\chi}^{e(t)}_+$. While this could be done, here we shall not discard negative frequency modes as they are also physically interesting, in particular when switching internal times in cosmology, see fig.\ \ref{fig:cs} (e.g., we would be discarding forward evolution in $\phi$). Therefore, we keep the normalization as in (\ref{identify2}), so that positive and negative frequency modes can be simultaneously normalized.

It is now easy to see that with an additional transformation for the measure
\ba
\widehat{\sqrt{|p_e|}}\,\ct_t\,\ket{\psi}_{\rm phys}=\f{1}{2}\,\ket{-\epsilon}_t \ket{\psi}^{e(t)}_++\f{1}{2}\,\ket{+\epsilon}_t\ket{\psi}^{e(t)}_-\,,\label{lasttrans}
\ea
we can identify $\ket{\psi}^{e(t)}_\pm$ with the reduced states (\ref{redstate}) on $\ch^{e(t)}_\pm$. We also recover the reduced evolving observables (\ref{Dirac4}) in the corresponding sectors (here the $+i$ term in (\ref{Dirac5}) is crucial)
\ba
\widehat{\sqrt{|p_e|}}\,\ct_t\,\hat{E}(\tau)\,\ct_t^{-1}\, \widehat{(\sqrt{|p_e|})^{-1}}&=&\left(|\hat{p}_e|\,\tau+\hat{E}\right)\,\theta(-\hat{p}_t)+\left(-|\hat{p}_e|\,\tau+\hat{E}\right)\,\theta(\hat{p}_t)\nn\\
&=&\hat{E}_+(\tau)\,\theta(-\hat{p}_t)+\hat{E}_-(\tau)\,\theta(\hat{p}_t)\,,\nn\\
\widehat{\sqrt{|p_e|}}\,\ct_t\,\hat{p}_e(\tau)\,\ct_t^{-1}\, \widehat{(\sqrt{|p_e|})^{-1}} &=&\hat{p}_e\,\theta(-\hat{p}_t)+\hat{p}_e\,\theta(\hat{p}_t)\,.\label{toatrans2}
\ea

Projecting onto the classical gauge fixing conditions $t=0$, in some analogy to the Page-Wootters construction \cite{Page:1983uc},  removes the redundant clock-slot and finally yields the states of the reduced theory
\ba
\ket{\psi}_{\pm}^{e(t)}=2\sqrt{2\pi}\,{}_t\bra{t=0}\,\theta(\mp\hat{p}_t)\,\widehat{\sqrt{|p_e|}}\,\ct_t\,\ket{\psi}_{\rm phys}\,.\label{q0state}
\ea
This projection is compatible with the observables and the inner product. Its image is the Heisenberg picture on $\ch_\pm^{e(t)}$; e.g., (\ref{q0state}) can be interpreted as an initial state at $t=0$. This completes the quantum reduction from the Dirac quantized theory to the reduced one relative to internal time $t$, see fig.\ \ref{fig:sum}.

\subsection{Quantum internal time switches}

This quantum reduction procedure now enables us to switch from the relational quantum dynamics relative to $t$ to that relative to $e$ \cite{Hoehn:2018aqt}. Just like in the classical case, we can thus interchange the roles of $t$ and $e$ and the following is the quantum analog of it. In analogy to a coordinate change on a manifold, we have to invert the quantum reduction map associated to $t$ and concatenate it with that associated to $e$. This will map from the reduced Hilbert spaces $\ch^{e(t)}_\pm$ via the internal-time-neutral $\ch_{\rm phys}$ to $\ch^{t(e)}_\pm$:
\ba
\hat{\cs}_{t_+\to e_\pm}:\ch_+^{e(t)}\rightarrow\ch_\pm^{t(e)}\,,\q\q\q\q\q\q\q\q\q\q\q\q\nn\\
\hat{\cs}_{t_+\to e_\pm}:=2\sqrt{2\pi}\,{}_e\bra{e=0}\,\theta(\mp\hat{p}_e)\,\widehat{\sqrt{|\hat{p}_t|}}\,\ct_{e\pm}\,\ct_{t+}^{-1}\, \widehat{(\sqrt{|p_e|})^{-1}}\,\f{\ket{p_t=-\epsilon}_t}{2}\otimes\,,\label{t2q}
\ea
where $\ct_e$ is identical to (\ref{trivialt}), except that $t$ and $e$ are interchanged.
Here, $\ket{p_t=-\epsilon}_t\otimes$ mean tensoring the input state $\ket{\psi}_+^{e(t)}$ with this factor, which amounts to restoring gauge invariance as $\ket{p_t=-\epsilon}_t=1/\sqrt{2\pi}\,\int\,\mathrm{d}t\,\exp(-i\,t\,\epsilon)\ket{t}_t$ averages over the classical gauge fixing conditions $t=const$. Similarly, for the negative frequency modes, we have
\ba
\hat{\cs}_{t_-\to e_\pm}:\ch_-^{e(t)}\rightarrow\ch_\pm^{t(e)}\,,\q\q\q\q\q\q\q\q\q\q\q\q\nn\\
\hat{\cs}_{t_-\to e_\pm}:=2\sqrt{2\pi}\,{}_e\bra{e=0}\,\theta(\mp\hat{p}_e)\,\widehat{\sqrt{|\hat{p}_t|}}\,\ct_{e\pm}\,\ct_{t-}^{-1}\, \widehat{(\sqrt{|p_e|})^{-1}}\,\f{\ket{p_t=+\epsilon}_t}{2}\otimes\,,\label{t2q2}
\ea

It is clear that, just as in the classical case, the internal time switches have to preserve the four quadrants of fig.\ \ref{fig:cs}. Indeed, in appendix \ref{app_qswitch} we show that
\ba
\hat{\cs}_{t_+\to e_\pm}\,\ket{\psi}_+^{e(t)}=
\theta(-\hat{p}_t)\,\ket{\psi}^{t(e)}_\pm\,,
\q\q\q\q \hat{\cs}_{t_-\to e_\pm}\,\ket{\psi}_-^{e(t)}=\theta(\hat{p}_t)\,\ket{\psi}^{t(e)}_\pm\,,\label{right}
\ea
where the reduced states on the left and right hand sides of the equations correspond via (\ref{identify2}) to the same physical state. We also demonstrate in appendix \ref{app_qswitch} that the complicated expressions (\ref{t2q}, \ref{t2q2}) vastly simplify, being equivalent to
\ba
\hat{\cs}_{t_+\to e_\pm}\equiv \cp_{e_\pm\to t_+}\,\theta(\mp\hat{p}_e)\,,\q\q\q\q\q\q \hat{\cs}_{t_-\to e_\pm}\equiv \cp_{e_\pm\to t_-}\,\theta(\mp\hat{p}_e)\,,\label{equiv}
\ea
where we have introduced the clock-switch operators 
\ba
\cp_{e_\pm\to t_+}\,\ket{p_e}_e:=\ket{-|p_e|\,}_t\,,\q\q\q\q\q \cp_{e_\pm\to t_-}\,\ket{p_e}_e:=\ket{\,|p_e|\,}_t\,\label{swap}
\ea
in close analogy to \cite{Hoehn:2018aqt} and the parity-swap operator of \cite{Giacomini:2017zju, Vanrietvelde:2018pgb}.

The quantum clock switch procedure can be summarized in a commutative diagram: 
\begin{center}
\begin{tikzcd}
& \mathcal{H}_{\textrm{phys}}  \arrow[rd, "\mathcal{T}_{e\pm}"]& \\
\mathcal{H}^{e(t)}_{\textrm{phys}} \arrow[ru, "\mathcal{T}_{t+}^{-1}"] & & \mathcal{H}^{t(e)}_{\textrm{phys}} \arrow[d, "2\sqrt{2\pi}\,{}_e\bra{e=0}\,\theta(\mp\hat{p}_e)\,\widehat{\sqrt{|\hat{p}_t|}}"]\\
\mathcal{H}_+^{e(t)} \arrow[rr, "\hat{\mathcal{S}}_{t_+ \to q \pm}"] \arrow[u, "\widehat{(\sqrt{|p_e|})^{-1}}\,\f{1}{2}\ket{p_t=-\epsilon}_t\otimes"]  & & \mathcal{H}^{t(e)}_{\pm}
\end{tikzcd}
\end{center}
and analogously for (\ref{t2q2}). Notice that the quantum clock switch thereby has the structure $\varphi_e\circ\varphi_t^{-1}$ of a coordinate transformation, where the internal-time-neutral $\ch_{\rm phys}$ assumes the role of the `manifold'. This is the appropriate structure for a quantum notion of general covariance that pertains to switching between the descriptions of the physics relative to different quantum reference systems, supporting the arguments in \cite{Vanrietvelde:2018pgb, Vanrietvelde:2018dit,Hoehn:2018aqt }. 

The inverse clock switch from $e$ to $t$ is due to the symmetry of the problem the same as above, except that one has to interchange the $e$ and $t$ labels everywhere. It is now easy to check how the elementary observables transform from $\ch_\pm^{e(t)}$ to $\ch_\pm^{t(e)}$:
\ba
\hat{\cs}_{t_\pm\to e_+}\,\hat{E}\,\hat{\cs}_{e_+\to t_\pm} &=& \hat{T}\,\theta(\mp\hat{p}_t)\,,\q\q\q\q \hat{\cs}_{t_\pm\to e_+}\,\hat{p}_e\,\hat{\cs}_{e_+\to t_\pm} =\pm \hat{p}_t\,\theta(\mp\hat{p}_t)\,,\nn\\
\hat{\cs}_{t_\pm\to e_-}\,\hat{E}\,\hat{\cs}_{e_-\to t_\pm} &=& \hat{T}\,\theta(\mp\hat{p}_t)\,,\q\q\q\q \hat{\cs}_{t_\pm\to e_-}\,\hat{p}_e\,\hat{\cs}_{e_-\to t_\pm} =\mp \hat{p}_t\,\theta(\mp\hat{p}_t)\,.\label{qswitch}
\ea
Notice that the image of $\hat{\cs}_{e_+\to t_\pm} $ is $\theta(-\hat{p}_e)\,\left(\ch_\pm^{e(t)}\right)$ so that in the first line one can set $\hat{p}_e=-|\hat{p}_e|$. Similarly, in the second line one can set $\hat{p}_e=|\hat{p}_e|$. Then it is obvious that the transformations (\ref{qswitch}) are exactly the quantum version of the classical maps between the corresponding reduced phase spaces in (\ref{clswitch}), which have been obtained through gauge transformations. While there are no gauge transformations in the quantum theory (except on $\ch_{\rm kin}$) \cite{Vanrietvelde:2018pgb, Vanrietvelde:2018dit,Hoehn:2018aqt }, this is their quantum analog.

These relations permit us to transform the reduced relational observables (\ref{Dirac4}) from $\ch_\pm^{e(t)}$ to $\ch_\pm^{t(e)}$:
\ba
\hat{\cs}_{t_\pm\to e_+}\,\hat{E}_\pm(\tau_t)\,\hat{\cs}_{e_+\to t_\pm} &=&\left(\pm|\hat{p}_t|\,\tau_t+ \hat{T}\right)\,\theta(\mp\hat{p}_t)\,,\nn\\
\hat{\cs}_{t_\pm\to e_-}\,\hat{E}_\pm(\tau_t)\,\hat{\cs}_{e_-\to t_\pm} &=& \left(\pm|\hat{p}_t|\,\tau_t+ \hat{T}\right)\,\theta(\mp\hat{p}_t)
\,\label{qswitch2}
\ea
($\hat{p}_{e_\pm}(\tau_t)$ is already transformed in (\ref{qswitch})). The right hand side is {\it not} $\hat{T}_\pm(\tau_e)$, despite looking like it, due to the appearance of $\tau_t$, which runs over the values of $t$, rather than $\tau_e$, which runs over the values of $e$. Instead, it is the representation of $\hat{E}_\pm(\tau_t)$ on $\ch_\pm^{t(e)}$ and could be used to set initial values $\tau_e^i$ for $e$ after the clock switch. 

In contrast to the classical case, there does not seem to be a unique procedure, given that $\hat{E}_\pm$ is now an operator. However, in analogy to the classical case, we can define the initial reading $\tau_e^i$ of the new clock $e$ in terms of expectation values, e.g.:
\ba
\tau_e^i:=\f{\la\hat{E}_\pm(\tau_t^f)\ra_\pm^{e(t)}}{\la\hat{p}_e\ra_\pm^{e(t)}}\,.\label{set}
\ea
Indeed, we prove in appendix \ref{app_qcontinuous} that this leads to exactly the classical `continuity' relation (\ref{ccontinuous}) in terms of expectation values
\ba
\Big\la\hat{T}_\pm(\tau_e^i)\Big\ra_\pm^{t(e)} = \tau_t^f\,\la\hat{p}_t\ra_\pm^{t(e)}\q\q\q\text{ on } \ch^{t(e)}_\pm\,,\label{qcontinuous}
\ea
so that one also finds a continuous quantum relational evolution, despite the intermediate clock switch.

\subsection{Illustration in concrete states}

\begin{figure}
    \centering
    \begin{subfigure}[b]{0.34\textwidth}
        \includegraphics[width=\textwidth]{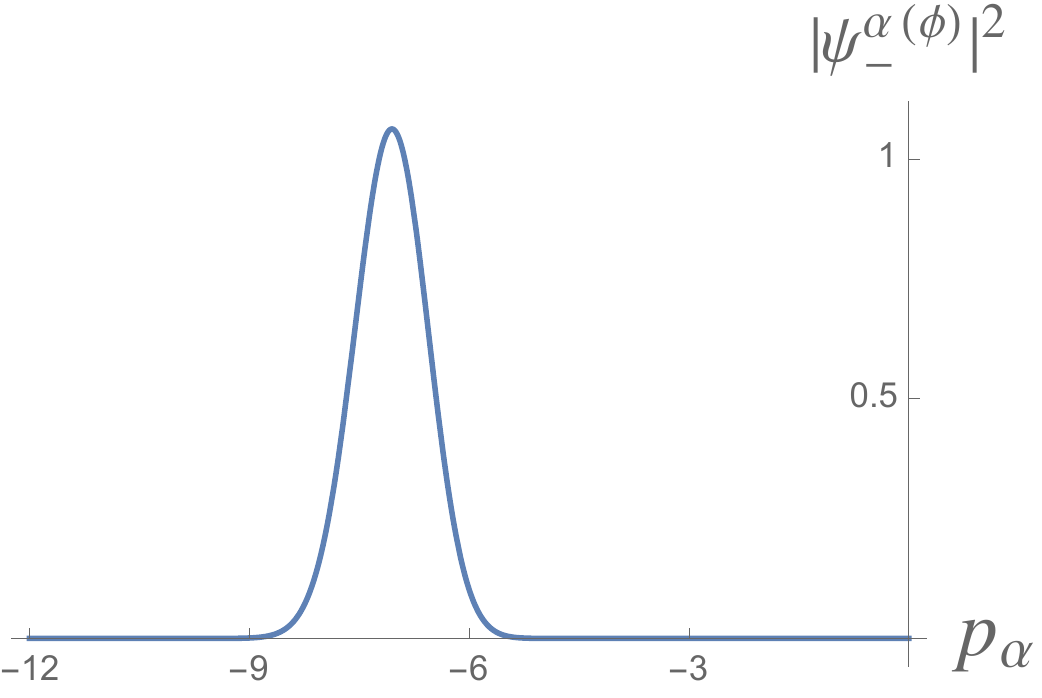}
        \caption{}
        \label{fig:1a}
    \end{subfigure}
 ~  \q\q\q\q\q\q\q\q\q
    \begin{subfigure}[b]{0.36\textwidth}
        \includegraphics[width=\textwidth]{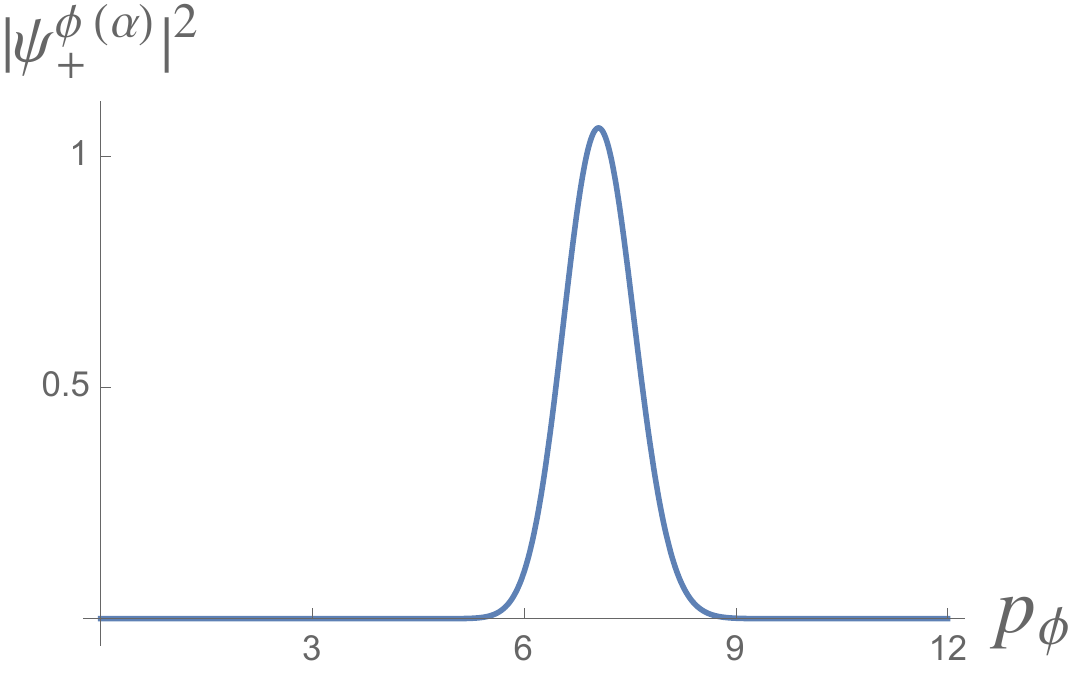}
        \caption{}
        \label{fig:1b}
    \end{subfigure}
    \caption{\small Reduced probability distributions coming from the same physical state (defined through (\ref{physstate}, \ref{n}) and here $n=100$), but described relative to the choices of (a) $\phi$ and (b) $\alpha$ as internal times in the `expanding-forward' (green) quadrant of fig.\ \ref{fig:cs}. Recall that in the reduced theory the usual modulus square of the wave function is the probability distribution, see (\ref{redIP}). Due to the symmetry of the model in $\alpha$ and $\phi$, reduced probability distributions will always behave symmetrically. }\label{fig:ex}
\end{figure}

\begin{SCfigure}
\centering
\includegraphics[width=.44\textwidth]{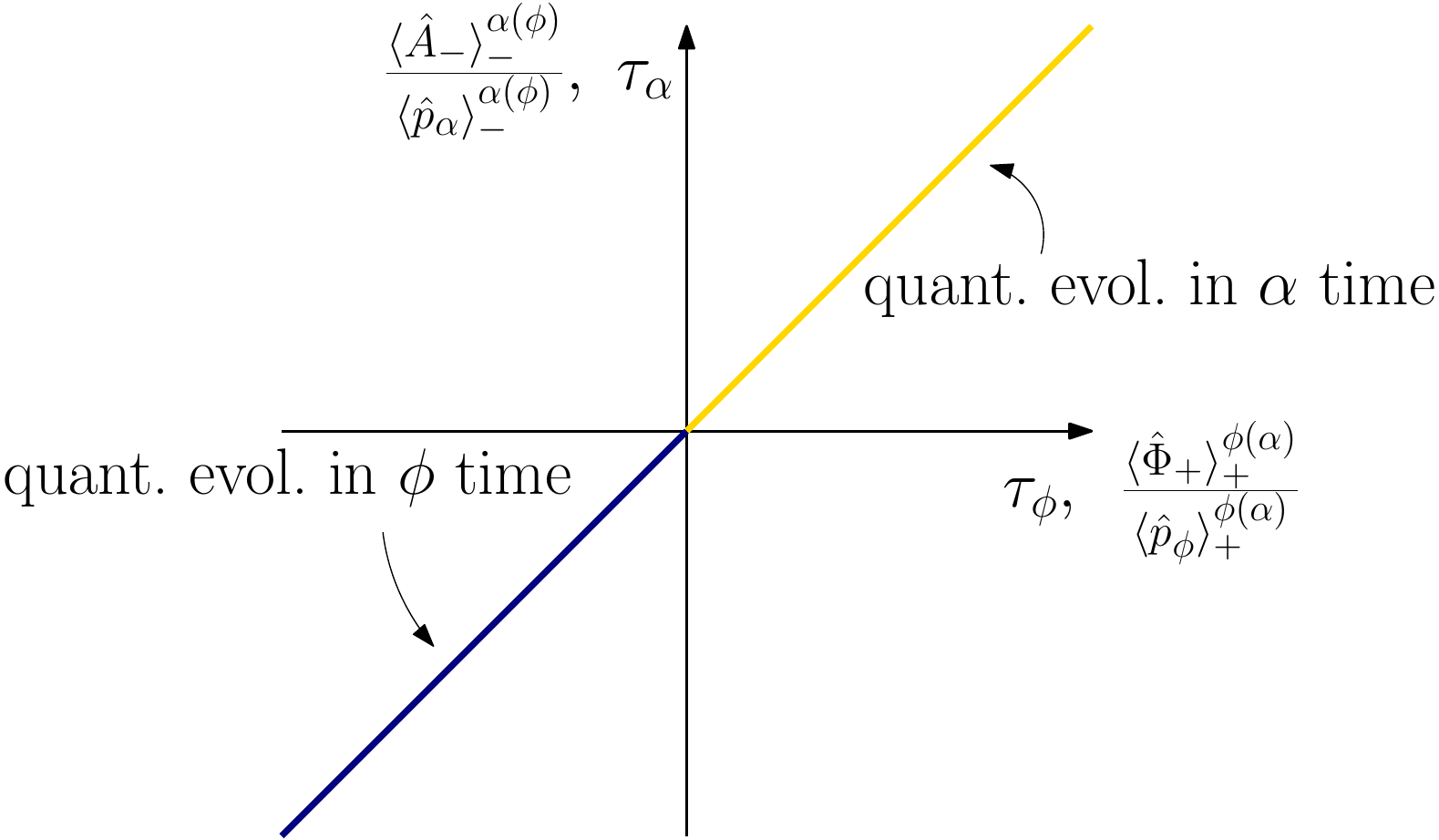}
        \caption{\small Illustration of the quantum relational evolution given in (\ref{ex1}, \ref{ex2}) for an internal time switch from $\phi$ to $\alpha$ at $\tau_\phi^f=\tau_\alpha^i=0$. The blue branch corresponds to the evolution of $\Big\la\hat{A}_-(\tau_\phi)\Big\ra_-^{\alpha(\phi)}/\la\hat{p}_\alpha\ra_-^{\alpha(\phi)}$ in $\tau_\phi$, while the golden branch depicts the evolution of $\Big\la\hat{\Phi}_+(\tau_\alpha)\Big\ra_+^{\phi(\alpha)}/\la\hat{p}_\phi\ra_+^{\phi(\alpha)}$ in $\tau_\alpha$. Together they trace out a continuous classical trajectory, describing an expanding universe.}
        \label{fig:switch}
    \end{SCfigure}

Let us briefly illustrate this internal time switch for example states. We pick semiclassical kinematical states, built according to the recipe for elliptic coherent states in \cite{ell} (and adapt the normalization):
\ba
\psi_{\rm kin}(p_\phi,p_\alpha)=\sqrt{\f{2}{\Gamma(n)}}\,(p_\phi+i\,p_\alpha)^n\,\exp\left(-\f{p_\alpha^2+p_\phi^2}{2}\right)\,.\label{n}
\ea
For concreteness, we restrict to the green quadrant in fig.\ \ref{fig:cs}, where we have $p_\alpha=-p_\phi\leq0$ and so an expanding universe with forward evolution in $\phi$. Using the Newton-Wigner type identification (\ref{identify2}), this gives semiclassical reduced negative and positive frequency wave functions on $\theta(-\hat{p}_\alpha)\big(\ch_-^{\alpha(\phi)}\big)$ and $\theta(\hat{p}_\phi)\big(\ch_+^{\phi(\alpha)}\big)$, respectively,
\ba
\psi_-^{\alpha(\phi)}(p_\alpha)=\sqrt{\f{2}{\Gamma(n)\,|p_\alpha|}}\,p_\alpha^n(i-1)^n\,\exp\left(-p_\alpha^2\right)\,,\q\q\q\q \psi_+^{\phi(\alpha)}(p_\phi)=\sqrt{\f{2}{\Gamma(n)\,p_\phi}}\,p_\phi^n(i-1)^n\,\exp\left(-p_\phi^2\right)\,.\nn
\ea
For visualization, we provide plots of their probability distributions in fig.\ \ref{fig:ex}.
One easily finds that $\la\hat{A}\ra_-^{\alpha(\phi)}=\la\hat{\Phi}\ra_+^{\phi(\alpha)}=0$, where $\hat{A}, \hat{\Phi}$ are the reduced quantizations (\ref{affrep}) of $A=\alpha\,p_\alpha$ and $\Phi=\phi\,p_\phi$, and
\ba
\Big\la\hat{A}_-(\tau_\phi)\Big\ra_-^{\alpha(\phi)}&=&\tau_\phi\,\la\hat{p}_\alpha\ra_-^{\alpha(\phi)}=-\tau_\phi\,\f{\Gamma(n+\f{1}{2})}{\sqrt{2}\,\Gamma(n)}\,,\nn\\
\Big\la\hat{\Phi}_+(\tau_\alpha)\Big\ra_+^{\phi(\alpha)}&=&\tau_\alpha\,\la\hat{p}_\phi\ra_+^{\phi(\alpha)}=+\tau_\alpha\,\f{\Gamma(n+\f{1}{2})}{\sqrt{2}\,\Gamma(n)}\,.\label{ex1}
\ea
Suppose we evolve first in $\phi$ and then switch to $\alpha$ time. Then invoking (\ref{set}) immediately yields 
\ba
\tau_\alpha^i=\tau_\phi^f\q\q\q\Rightarrow\q\q\q \Big\la\hat{\Phi}_+(\tau^i_\alpha)\Big\ra_+^{\phi(\alpha)}&=&\tau_\phi^f\,\la\hat{p}_\phi\ra_+^{\phi(\alpha)}\,.\label{ex2}
\ea
This simple switch from $\phi$ to $\alpha$ time is illustrated in fig.\ \ref{fig:switch}.

%

\section{Perspective on the `wave function of the universe'}\label{sec_conc}

We have illustrated in a very simple quantum cosmological model, namely the flat FRW universe with massless scalar field, how to consistently switch between the quantum relational dynamics relative to the scale factor and that relative to the field used as internal times. In particular, just like in the classical case, the quantum relational evolution is {\it continuous}, despite the intermediate internal time switch, and no information gets lost. This extends the quantum clock switch method of \cite{Hoehn:2018aqt} (see also \cite{Vanrietvelde:2018pgb, Vanrietvelde:2018dit,Giacomini:2017zju} for spatial reference systems) to the relativistic case and offers a full Hilbert space alternative to the semiclassical effective approach of \cite{Bojowald:2010xp,Bojowald:2010qw, Hohn:2011us}.

Owing to the symmetry of the model in $\phi$ and $\alpha$, the internal time switches are particularly simple here and the relational dynamics in a given physical (i.e.\ internal-time-neutral) state looks essentially `the same' relative to these two possible choices (up to relabeling the evolving variables). This will no longer be the case in models which are not symmetric relative to different internal time choices, e.g.\ see \cite{Hoehn:2018aqt}, and especially not in the presence of the so-called global problem of time \cite{Kuchar:1991qf,Isham:1992ms,Anderson:2017jij,Bojowald:2010xp,Bojowald:2010qw, Hohn:2011us,Dittrich:2016hvj,Dittrich:2015vfa,Hajicek:1986ky, Hajicek:1994py,Hajicek:1995en,Rovelli:1990jm}, which arises, e.g., for interactions between evolving and internal time degrees of freedom  \cite{Hohn:2011us,Marolf:1994nz,Marolf:2009wp, Giddings:2005id}. However, our method is general and applies to generic models if one suitably takes into account the Gribov problem and the fact that a description relative to a choice of reference system, just like a coordinate choice, will generally not be globally valid \cite{Vanrietvelde:2018pgb, Vanrietvelde:2018dit,Hoehn:2018aqt ,Bojowald:2010xp,Bojowald:2010qw, Hohn:2011us}.

Indeed, the internal time switch proceeds in complete analogy to coordinate changes $\varphi_t\circ\varphi_{t'}^{-1}$ on a manifold \cite{Hoehn:2018aqt,Vanrietvelde:2018pgb }: it inverts the quantum reduction map relative to one time choice, mapping the corresponding reduced quantized theory back into the internal-time-neutral physical Hilbert space of the Dirac quantization and subsequently applies the quantum reduction map to the reduced quantization relative to the other internal time choice. The same compositional structure appears for changes of spatial quantum reference systems \cite{Vanrietvelde:2018pgb, Vanrietvelde:2018dit}. 
This permits us to interpret the physical Hilbert space of the Dirac quantized theory as encoding the `perspective-neutral' (i.e.\ reference-system-neutral) physics \cite{Hoehn:2017gst,Vanrietvelde:2018pgb, Vanrietvelde:2018dit,Hoehn:2018aqt} and the quantum reduction maps as defining `quantum coordinate' descriptions of these physics relative to a choice of quantum reference system. This is precisely the structure that one would expect for establishing a genuine {quantum} notion of general covariance, which refers to the ability to consistently switch, within {one} theory, between arbitrary choices of quantum reference systems, each of which can be used as a vantage point to describe the physics of the remaining degrees of freedom.

Accordingly, in line with our earlier discussion in \cite{Vanrietvelde:2018pgb, Vanrietvelde:2018dit,Hoehn:2018aqt}, we thus propose to define a {\it complete} relational quantum theory, admitting a quantum general covariance, as the {\it conjunction} of the quantum-reference-system-neutral Dirac quantized theory and the multitude of reduced quantum theories associated to the different choices of quantum reference system. Just like the classical theory contains a (perspective-neutral) constraint surface {and} the multitude of reduced phase spaces, together comprising a complete classical description, the complete quantum theory contains their corresponding quantum structures, as illustrated here, and this is a complete quantum description. Specifically, we propose this conjunction to overcome the so-called multiple choice facet of the problem of time \cite{Kuchar:1991qf,Isham:1992ms} (the arguments of which could also be applied to spatial reference systems) and to turn it into a multiple choice {\it feature} of the complete relational quantum theory \cite{Hoehn:2018aqt}.

For simplicity, we have illustrated the novel procedure using the Wheeler-DeWitt approach in the Dirac quantization, however, the loop quantization of the simple model of this article can actually be formulated in the same physical Hilbert space \cite{Ashtekar:2007em}. It is thus suggestive that the present framework for switching internal times can be extended to loop quantum cosmology as well. To this end, however, at least two loop quantization related subtleties need be suitably taken into account. For instance, loop quantization leads to superselection sectors in the geometric degrees of freedom \cite{Banerjee:2011qu}. Presumably, the framework should be applied per superselection sector, although subtleties remain to be checked as different sectors might have different physical properties. Secondly, loop quantization leads to a deformation of gauge covariance as embodied in the constraint algebra \cite{Bojowald:2012ux,Bojowald:2015zha,Bojowald:2015sta}. While this should not be a problem for homogeneous cosmological models, it arises as an additional challenge when attempting to extend the framework to a loop quantization of inhomogeneous models where non-trivial diffeomorphism constraints arise. This would be relevant, e.g., when studying relational dynamics in the context of loop quantum cosmology modifications of the 'no-boundary' proposal \cite{Bojowald:2018gdt}.

Our proposal also entails a novel perspective on the `wave function of the universe', i.e.\ the {global} quantum state for the universe as a whole, which appears ubiquitously and in various interpretational guises in quantum cosmology \cite{Everett:1956dja, Everett:1957hd,Hartle:1983ai,Hawking:1983hj,Hawking:1983hn,page2,Bojowald:2011zzb,martinbuch,Ashtekar:2011ni,Banerjee:2011qu}. It is usually taken to be a solution to the Wheeler-DeWitt equation, in the present article (\ref{qCH}), and thereby a physical state of the Dirac quantized theory. The proposal here suggests to view the `wave function of the universe' as a perspective-neutral global state that thereby does {\it not} admit an immediate physical interpretation; it is not the description of the universe relative to any physical reference system. Instead, while in the simple model here we have only illustrated it for two possible choices, it contains the information about {\it all} the relative states at once, i.e.\ all the  descriptions relative to all possible choices of quantum reference system, and provides the crucial linking structure between all these relative descriptions. In fact, it is these relative reduced states that admit the immediate physical interpretation and should be taken as relevant for observational and operational predictions (although the `wave function of the universe' encodes that information too).

This offers a consistent link between operationally significant subsystem structures in quantum cosmology and gravity, relative to a choice of quantum reference system, and a perspective-neutral (in particular, observer-independent) global state that contains all degrees of freedom \cite{Hoehn:2017gst}. Specifically, this also suggests a novel perspective on the notorious problem of how to interpret the  probabilities defined by the `wave function of the universe'. While the global probability density defined by it through the physical inner product (here (\ref{PIP})) does not admit an immediate operational interpretation, the `wave function of the universe' gives rise to all the relative states through quantum reduction, and these {\it do} admit an immediate physical interpretation. Indeed, the relative states admit a physically relevant reduced probability distribution (here via (\ref{redIP})), and the quantum reduction always implies their inner product through the inner product of the corresponding physical states (here see (\ref{inp}) and \cite{Vanrietvelde:2018pgb, Vanrietvelde:2018dit,Hoehn:2018aqt} for further examples of the method). However, crucially, the two kinds of probability distributions live on different spaces: the `wave function of the universe' technically defines an abstract probability distribution over {\it all} the degrees of freedom of the universe, while the relative states define a probability distribution over all degrees of freedom of the universe, except those of the associated reference system. As such, the latter admits the interpretation as the probability distribution `seen' by that reference system.

Note that the proposal here is general and not specific to any detailed interpretation of quantum theory and its probabilities. There is no obvious reason why it should conflict with any of many-worlds, relational, QBism, consistent histories, Copenhagen, or realist interpretations. In particular, it is worthwhile to point out that it might 
actually reconcile relational and informational  state interpretations \cite{Rovelli:1995fv,rovelli2018space,Hoehn:2014uua,Hoehn:2015zom,Hoehn:2016otu,Martin-Dussaud:2018kmh, koberinski_muller_2018, brukner2017quantum,Fuchs:2016qml} with the global `wave function of the universe'. While the details depend on the specific interpretation, relational interpretations take a state to be defined relative to an agent, or, more generally, reference system, and this state is taken to be the observer's `catalog of knowledge' about the observed system. One can then argue \cite{Hoehn:2017gst,Hoehn:2014uua } that such interpretations deny a global operationally meaningful quantum state as the self-reference problem \cite{breuer1995impossibility,dalla1977logical} impedes a given observer or reference system to infer the global state of the entire universe (incl.\ itself) from its interactions with the rest.  Accordingly, relative to any subsystem, one can assign a `catalog of knowledge' about the rest of the universe but, without external observer or reference frame, there can then be no global, operationally meaningful `catalog of knowledge' about the entire universe at once (see also related discussions in \cite{Crane:1995qj,Markopoulou:2002ru,Markopoulou:2007qg,Hackl:2014txa }). In the proposal of this article, the global `wave function of the universe' indeed does not admit an immediate operational interpretation as an informational state, yet it links all the different relational reference system perspectives on the universe consistently \cite{Hoehn:2017gst}, something that was missing, e.g., in the discussion of \cite{Rovelli:1995fv,rovelli2018space,Hoehn:2014uua,Crane:1995qj,Markopoulou:2002ru,Markopoulou:2007qg, brukner2017quantum,Fuchs:2016qml}. 

Specifically, this might reconcile the seemingly subjective relational states (an observer's `catalog of knowledge') with the objective `wave function of the universe'. Being a physical system too, the subjective degrees of beliefs, i.e.\ `catalogs of knowledge' of any observer about states of other systems should be encoded in physical degrees of freedom of this observer. But the `wave function of the universe' -- as a perspective-neutral global state -- encodes {\it all} physical degrees of freedom of the universe and thus `knows', in particular, what information any observing system has in its memory. Hence, while the relative states may be interpreted as subjective `catalogs of knowledge' of the observing systems, the `wave function of the universe', as proposed here, contains all these `catalogs of knowledge' at once and would actually consistently and objectively link them. However, to manifest this more specific interpretation, one would have to clarify how state collapses occur in the relative descriptions from measurement interactions at the perspective-neutral level. That is, one has to revisit the measurement problem (and specifically the Wigner friend paradox \cite{wigner1995remarks,deutsch1985quantum,Rovelli:1995fv,brukner2017quantum,frauchiger2016single}), but now with a complete relational quantum theory at hand, as proposed here, which contains both a perspective-neutral description and all the individual perspectives, a structure that was not available before.

Finally, it can be shown in simple examples that quantum correlations will generally depend on the choice of quantum reference system \cite{Vanrietvelde:2018pgb,Giacomini:2017zju}. This immediately raises some interesting questions since both quantum reference systems and quantum correlations appear ubiquitously in quantum cosmology. For example, given the phenomenological importance of CMB correlations and propagators, does the quantum frame dependence of correlations, which surely has to be expected in quantum cosmology too, have {\it any} observational significance? This question could be studied, e.g., in Bianchi models with inhomogeneous perturbations and the tools for these investigations are now, in principle, available.

\section*{Acknowledgements}
I thank Bianca Dittrich, Steffen Gielen and Merce Mart\'in-Benito for discussion. The project leading to this publication has initially received funding from the European Union's Horizon 2020 research
and innovation programme under the Marie Sklodowska-Curie grant agreement No 657661. The author also acknowledges support through a Vienna Center for Quantum Science and Technology Fellowship.

\appendix

\section{Hermiticity of the relational observable $\hat{E}(\tau)$ on $\ch_{\rm phys}$}\label{app_SA}

We prove the claim that $\hat{E}(\tau)$ as given in (\ref{Dirac5}) is Hermitian with respect to the physical inner product. To this end, it suffices to consider the symmetric quantization of the Dirac observable $L$ in (\ref{L}) on $\ch_{\rm kin}$
\ba
\hat{L}=\f{1}{2}(\hat{p}_\phi\,\hat{\phi}+\hat{\phi}\,\hat{p}_\phi+\hat{p}_\alpha\,\hat{\alpha}+\hat{\alpha}\,\hat{p}_\alpha)\,.
\ea
This is a Hermitian and, in particular, self-adjoint operator on $\ch_{\rm kin}$. However, it fails to be Hermitian with respect to the physical inner product. To see this, note that
\ba
[\hat{C}_H,\hat{L}\,]=-2i\,\hat{C}_H\,.\label{chl}
\ea
Hence, $\hat{L}$ commutes with the constraint {\it only} on $\ch_{\rm phys}$. The physical inner product (\ref{PIP}) comes from group averaging \cite{Hartle:1997dc,Marolf:1995cn,Marolf:2000iq,Thiemann:2007zz,Hoehn:2018aqt,Ashtekar:1995zh} and is given by
\ba
\la\psi|\chi\ra_{\rm phys}:=\la \psi_{\rm kin}|\,\delta(\hat{C}_H)\,|\chi_{\rm kin}\ra\,,\label{PIP2}
\ea
where $\la\cdot|\cdot\ra$ is the standard inner product on $\ch_{\rm kin}$ and 
\ba
\delta(\hat{C}_H)=\f{1}{2\pi}\,\int_{-\infty}^{+\infty}\,\mathrm{d}s\,e^{i\,s\,\hat{C}_H}\,.
\ea
Using (\ref{chl}) one easily finds $[\hat{C}_H^n,\hat{L}]=-2i\,n\,\hat{C}_H^n$ and thereby
\ba
[\delta(\hat{C}_H),\hat{L}\,]&=&\f{1}{\pi}\,\int\,\mathrm{d}s\,s\,\hat{C}_H \,e^{i\,s\,\hat{C}_H}=-2i\,\f{\mathrm{d}}{\mathrm{d}x}\f{1}{2\pi}\,\int\,\mathrm{d}s e^{i\,x\,s\,\hat{C}_H}\Big|_{x=1}\nn\\
&=&-2i\,\f{\mathrm{d}}{\mathrm{d}x}\,\delta(x\,\hat{C}_H)\Big|_{x=1}=-2i \,\f{\mathrm{d}}{\mathrm{d}x}\,|x|^{-1}\Big|_{x=1}\,\delta(\hat{C}_H)=2i\,\delta(\hat{C}_H)\,.\label{delch}
\ea
From this result it is clear that $\hat{L}$ is {\it not} Hermitian with respect to (\ref{PIP2}). However, using (\ref{delch}), we have
\ba
\la \psi_{\rm kin}|\,(\hat{L}+i)\,\delta(\hat{C}_H)\,|\chi_{\rm kin}\ra=\la \psi_{\rm kin}|\,\delta(\hat{C}_H)\,(\hat{L}-i)\,|\chi_{\rm kin}\ra=\la (\hat{L}+i)\,\delta(\hat{C}_H)\,\psi_{\rm kin}\,|\chi_{\rm kin}\ra\,,
\ea
where in the last step we have made use of the fact that both $\hat{L}$ and $\delta(\hat{C}_H)$ are symmetric on $\ch_{\rm kin}$. Consequently, $\hat{L}+i$ {\it is} Hermitian with respect to the physical inner product and, in turn, also $\hat{E}(\tau)$ in (\ref{Dirac5}). This operator can also be densely defined and is thus essentially self-adjoint.

\section{Changes of internal times in the quantum theory}\label{app_qswitch}

We begin by proving the left equation in (\ref{right}). Recall that
\ba
\hat{\cs}_{t_+\to e_\pm}:=2\sqrt{2\pi}\,{}_e\bra{e=0}\,\theta(\mp\hat{p}_e)\,\widehat{\sqrt{|\hat{p}_t|}}\,\ct_{e\pm}\,\ct_{t+}^{-1}\, \widehat{(\sqrt{|p_e|})^{-1}}\,\f{\ket{p_t=-\epsilon}_t}{2}\otimes\,.
\ea
We make use of the definition of the reduced positive and negative frequency wave functions (\ref{identify2}) and
\ba
\psi_{\rm kin}^{t(e)}\Big(p_e=\mp|p_t|,|p_t|\Big)&=&\psi_{\rm kin}^{e(t)}\Big(|p_e|,\mp|p_e|\Big)\,,\nn\\
 \psi_{\rm kin}^{t(e)}\Big(p_e=\mp|p_t|,-|p_t|\Big)&=&\psi_{\rm kin}^{e(t)}\Big(-|p_e|,\mp|p_e|\Big)\,,\label{id3}
  \ea
which is implied by (\ref{compact}). Then,
\ba
\hat{\cs}_{t_+\to e_\pm}\,\ket{\psi}_+^{e(t)}&=&\hat{\cs}_{t_+\to e_\pm}\,\int_{-\infty}^\infty\,\mathrm{d}p_e\,\f{\psi_{\rm kin}^{e(t)}(-|p_e|,p_e)}{\sqrt{|p_e|}}\,\ket{p_e}_e\nn\\
&=&2\sqrt{2\pi}\,{}_e\bra{e=0}\,\theta(\mp\hat{p}_e)\,\widehat{\sqrt{|\hat{p}_t|}}\,\ct_{e\pm}\,\int_{-\infty}^\infty\,\f{\mathrm{d}p_e}{2|p_e|}\,\psi_{\rm kin}^{e(t)}(-|p_e|,p_e)\,\ket{-|p_e|\,}_t\ket{p_e}_e\nn\\
&=& 2\sqrt{2\pi}\,{}_e\bra{e=0}\,\theta(\mp\hat{p}_e)\,\widehat{\sqrt{|\hat{p}_t|}}\,\ct_{e\pm}\,\nn\\
&&\q\q\q\times\int_{-\infty}^0\,\f{\mathrm{d}p_t}{2|p_t|}\,\Big[\psi_{\rm kin}^{t(e)}(-|p_t|,p_t)\,\ket{-|p_t|\,}_e\ket{p_t}_t+\psi_{\rm kin}^{t(e)}(|p_t|,p_t)\,\ket{\,|p_t|\,}_e\ket{p_t}_t\Big]\nn\\
&=&\sqrt{2\pi}\,{}_e\bra{e=0}\,\theta(\mp\hat{p}_e)\int_{-\infty}^0\,\mathrm{d}p_t\,\Big[\psi_{+}^{t(e)}(p_t)\,\ket{-\epsilon}_e\ket{p_t}_t+\psi_{-}^{t(e)}(p_t)\,\ket{+\epsilon}_e\ket{p_t}_t\Big]\nn\\
&=&\theta(-\hat{p}_t)\,\ket{\psi}_\pm^{t(e)}\,.\nn
\ea
From the second to the third line, we have performed a variable change $p_e=p_t$ for $p_e<0$ and $p_e=-p_t$ for $p_e>0$ and used (\ref{id3}). 

To prove that this transformation is equivalent to $\cp_{e_\pm\to t_+}\,\theta(\mp\hat{p}_e)$, as claimed in (\ref{equiv}), where $\cp_{e_\pm\to t_+}$ is defined in (\ref{swap}), write
\ba
\ket{\psi}_+^{e(t)}=\int_{-\infty}^0\,\f{\mathrm{d}p_e}{\sqrt{|p_e|}}\,\psi^{e(t)}_{\rm kin}(-|p_e|,p_e)\,\ket{p_e}_e+\int_{0}^{+\infty}\,\f{\mathrm{d}p_e}{\sqrt{|p_e|}}\,\psi^{e(t)}_{\rm kin}(-|p_e|,p_e)\,\ket{p_e}_e\,,\nn
\ea
perform a variable transformation $p_e=p_t$ in the left and $p_e=-p_t$ in the right integral and invoke (\ref{id3}) and the definition (\ref{swap}).

The right equations in (\ref{right}, \ref{equiv}) are shown in complete analogy.

\section{Continuity of the quantum relational dynamics during a  switch}\label{app_qcontinuous}

We briefly prove the continuity of the quantum relational dynamics, as expressed in (\ref{qcontinuous}), notwithstanding the intermediate internal time switch. For concreteness, we restrict our attention to one quadrant of fig.\ \ref{fig:cs}, e.g.\ either the green or red quadrant where $p_\alpha=-p_\phi$. Notice first that (\ref{equiv}) implies
\ba
 \begin{tikzcd}[every arrow/.append style={shift left}]
\theta(-\hat{p}_e)\big(\ch_-^{e(t)}\big) \arrow{r}{\hat{\cs}_{t_-\to e_+}} &[4em] \theta(+\hat{p}_t)\big(\ch_+^{t(e)}) \arrow{l}{\tiny\hat{\cs}_{e_+\to t_-}}\,.
 \end{tikzcd}
\ea
Clearly, using (\ref{Dirac4}), we have
\ba
\Big\la\hat{E}_-(\tau_t)\Big\ra_-^{e(t)}&=&-\tau_t\,\la|\hat{p}_e|\ra_-^{e(t)}+\la\hat{E}\ra_-^{e(t)}= \tau_t\,\la\hat{p}_e\ra_-^{e(t)}+\la\hat{E}\ra_-^{e(t)} \q\q\text{ on }\theta(-\hat{p}_e)\big(\ch_-^{e(t)}\big) \,,\nn\\
\Big\la\hat{T}_+(\tau_e)\Big\ra_+^{t(e)}&=&+\tau_e\,\la|\hat{p}_t|\ra_+^{t(e)}+\la\hat{T}\ra_+^{t(e)}= \tau_e\,\la\hat{p}_t\ra_+^{t(e)}+\la\hat{T}\ra_+^{t(e)}  \,\q\q\text{ on }\theta(+\hat{p}_t)\big(\ch_+^{t(e)})\,.\nn
\ea
Setting now the initial value of the new clock $e$, as in (\ref{set}), to 
\ba
\tau_e^i:=\f{\la\hat{E}_-(\tau_t^f)\ra_-^{e(t)}}{\la\hat{p}_e\ra_-^{e(t)}}= \tau_t^f+\f{\la\hat{E}\ra_-^{e(t)}}{\la\hat{p}_e\ra_-^{e(t)}}\,,
\ea
we find
\ba
\Big\la\hat{T}_+(\tau_e^i)\Big\ra_+^{t(e)} = \tau_t^f\,\la\hat{p}_t\ra_+^{t(e)}+\f{\la\hat{E}\ra_-^{e(t)}}{\la\hat{p}_e\ra_-^{e(t)}}\,\la\hat{p}_t\ra_+^{t(e)}+ \la\hat{T}\ra_+^{t(e)} \,.\label{qcontinuous2}
\ea
Now we invoke (\ref{qswitch}) and, in particular,
\ba
\hat{\cs}_{t_-\to e_+}\,\hat{E}\,\hat{\cs}_{e_+\to t_-} &=& \hat{T}\,\theta(+\hat{p}_t)\,,\q\q\q\q \hat{\cs}_{t_-\to e_+}\,\hat{p}_e\,\hat{\cs}_{e_+\to t_-} =- \hat{p}_t\,\theta(+\hat{p}_t)\,.
\ea
Using (\ref{redIP}, \ref{right}), this implies
\ba
\f{\la\hat{E}\ra_-^{e(t)}}{\la\hat{p}_e\ra_-^{e(t)}}=-\f{\la\hat{T}\ra_+^{t(e)}}{\la\hat{p}_t\ra_+^{t(e)}}
\ea
and thereby
\ba
\Big\la\hat{T}_+(\tau_e^i)\Big\ra_+^{t(e)} = \tau_t^f\,\la\hat{p}_t\ra_+^{t(e)}\,,
\ea
as claimed. The proof for the other quadrants is completely analogous.

\bibliography{bibliography}{}
\bibliographystyle{utphys}

\end{document}